\UseRawInputEncoding
\documentclass[aps,prd,twocolumn,showpacs,preprintnumbers,superscriptaddress,nofootinbib,floatfix]{revtex4-2}
\usepackage{amsmath,amssymb,amsfonts}
\usepackage{bm}
\usepackage{bbm}
\usepackage{slashed}
\usepackage{graphicx}
\usepackage{color}
\usepackage{lineno}
\usepackage{comment}

\graphicspath{{figures/}}

\newcommand{\Tr}{\mathrm{Tr}}

\begin{document}
\title{Interplay of Scalar and Pseudo-Scalar Mesonic Channels in $\Lambda\bar{\Lambda}$ Spin Correlations}
\author{Kun Xu$^{a}$}
\affiliation{$^{a}$Beijing Institute of Technology, Beijing 100080, China}
\thanks{xukun@bit.edu.cn}
\begin{abstract}
The spin correlation of $\Lambda\bar{\Lambda}$ hyperon pairs produced in heavy-ion collisions is investigated through scalar and pseudo-scalar mesonic channels that mediate the creation of $s\bar{s}$ pairs. Because a $J=0$ meson forces $L=S$ while parity requires $(-1)^{L+1}=\pm1$, the two vertices project the pair onto distinct total-spin states: a scalar ($0^+$) vertex selects the $L=S=1$ configuration with a spin weight of $+1$, whereas a pseudo-scalar ($0^-$) vertex selects the $L=S=0$ configuration with a weight of $-3$; these weights are the traces of the correlation tensor, corresponding to $P_{s\bar{s}}=+1/3$ and $-1$, respectively. Implementing this framework within the three-flavor Nambu–Jona-Lasinio model at the Random-Phase-Approximation level, we compute the scalar and pseudo-scalar spectral weights. Our results show that the pseudo-scalar channel dominates, resulting in a negative net correlation fraction throughout. And the spin correlation also depends on the opening angle: strongest spin correlation at around $\theta=0$ while weakest at $\theta=\pi$. Furthermore, we show that the collision-energy dependence of the momentum-integrated spin correlation along the freeze-out line is sensitive to the trajectory’s path relative to the chiral phase boundary. Trajectories passing near the crossover region exhibit a non-monotonic structure, providing a sensitive probe for the location of the chiral phase transition.

\end{abstract}

\maketitle
\section{Introduction}
\label{sec:intro}
Non-central heavy-ion collisions(HIC) have been widely employed to explore the most subatomic vortical fluid via global hyperon polarization~\cite{STAR:2017ckg,STAR:2021beb}, as well as heavier hadrons with strange quark $\Xi^{-}$ and $\Omega^{-}$~\cite{STAR:2020xbm}.  The observed global spin polarization of (anti-)hyperons can be understood by the coupling between the fluid's kinematic vorticity and the particle's spin while magnetic field can induce their splitting~\cite{Becattini:2013fla,Karpenko:2016jyx,Xu:2022hql}. This phenomenon has opened a new window into the complex hydrodynamic evolution of strongly interacting medium created in HIC~\cite{Huang:2020vsp}.

Beyond the spin polarization of single fermions, the spin alignment of vector mesons has also been measured~\cite{ALICE:2019aid,STAR:2022fan}. It has been found theoretically that the global rotation leads to $\rho_{00}<1/3$~\cite{Wei:2023pdf} while the magnetic field induces $\rho_{00}>1/3$~\cite{Sheng:2022ssp}. However, the measured  global spin polarization of $\Lambda/\bar{\Lambda}$ gives $\rho_{00}\approx \frac{1}{3}-\frac{4}{9}P_{\Lambda}P_{\bar{\Lambda}}$ with $\frac{4}{9}P_{\Lambda}P_{\bar{\Lambda}} \approx 10^{-6}$, which is too small compared to the experimental data $\sim 10^{-2}$. $K^{*}$, however, does not show significant deviation from $\rho_{00}=1/3$.  On the other hand, a small but significant transverse polarization of inclusive $J/\psi$, corresponding to $\rho_{00}\approx0.25<1/3$, was also reported by ALICE~\cite{ALICE:2022dyy}. Thus, the global rotation or magnetic field in heavy ion collision is not enough to explain all these phenomena~\cite{Yang:2017sdk,Xia:2020tyd}.

Such discrepancies between theoretical models and local polarization measurements lead to theoretical proposals that local strong-force field fluctuations (e.g., the $\phi$ vector field) play a dominant role~\cite{Sheng:2019kmk,Sheng:2022sva,Sheng:2023urn,Sheng:2025puj}, and turbulent colour fields provide one concrete microscopic realization of such fluctuations~\cite{Muller:2022tcv}, as reviewed in Ref.~\cite{Chen:2023sfa}. In Ref.~\cite{Xu:2024kdh}, it was proposed that the spin correlation of quark pairs in HICs can be generated by local spin density fluctuations.  All examples show that the spin degrees of freedom of final-state hadrons retain a memory of the interactions that produced them, which indicates the spin correlation of fermions is the source of spin alignment of vector mesons.


%

Besides, there are more and more experiment data and theoretical proposals about spin correlation. For example, the spin correlation of $\Lambda-\bar{\Lambda}$ in a proton-proton collision has been also observed~\cite{STAR:2025njp},  which is believed to be inherited from spin-correlated strange quark-antiquark virtual pairs from QCD vacuum. In theory,  it was recently suggested by Chen~\cite{Chen:2024hki} that quark spin correlations can be significantly enhanced by the chiral phase transition. Consequently, spin correlation has emerged as a novel and sensitive probe for the properties of the QGP, such as QCD confinement and chiral phase transition.


The spin correlation between a $\Lambda$ and a $\bar{\Lambda}$ hyperon is extracted experimentally from the joint angular distribution of their weak-decay products~\cite{STAR:2025njp}. Boosting the decay (anti-)protons into the rest frames of their parent hyperons, the opening angle $\theta^*$ between the two momentum directions follows
\begin{align}
\frac{1}{N}\frac{dN}{d\cos\theta^*}=\frac{1}{2}\Big[1+\alpha_{\Lambda}\alpha_{\bar{\Lambda}}\,P_{\Lambda\bar{\Lambda}}\cos\theta^*\Big],
\label{eq:dist}
\end{align}
where $\alpha_{\Lambda,\bar{\Lambda}}$ are the weak decay asymmetry parameters and $P_{\Lambda\bar{\Lambda}}$ is the spin-correlation observable. 

Then it is natural to ask how the spins of a strange pair are correlated. In HIC, strange-quark pairs can be thermally produced in the QGP by gluon fusion $gg\to s\bar{s}$ or light-quark annihilation $q\bar{q}\to s\bar{s}$. On the other hand, the spin of strange-quark pairs can also be correlated by meson decay $M\rightarrow s\bar{s}$ at finite temperature, where production and decay of mesons have  reached equilibrium. Then the pair is subsequently converted into the observed $\Lambda\bar{\Lambda}$ pair, however, to simplify the calculations, we assume that the spin correlation of $\Lambda\bar{\Lambda}$ pairs is solely inherited from the $s\bar{s}$ pair~\cite{Lv:2024uev,Oliwa:2026qsc}, i.e., $P_{\Lambda\bar{\Lambda}} = P_{s\bar{s}}$.

In this paper we study the spin correlation of $s\bar{s}$ pairs produced from meson decay, specifically, from the scalar $\sigma$--$f_{0}$ channel and the pseudo-scalar $\eta$--$\eta'$ channel. Because a $J=0$ meson decays into a fermion--antifermion pair with $L=S$, and the pair parity is $(-1)^{L+1}$, the two channels are forced onto different total-spin states: the scalar vertex produces the pair in the $L=S=1$ (${}^{3}P_{0}$) configuration, the pseudo-scalar vertex in the $L=S=0$ (${}^{1}S_{0}$) configuration. Which of the two channels dominates therefore decides the sign and the magnitude of the hyperon spin correlation.  

On the OTHER hand, the mesons are presented by spectral function, which is obtained in the three-flavor Nambu--Jona-Lasinio model~\cite{Nambu:1961tp,Klevansky:1992qe,Hatsuda:1994pi} at the random-phase-approximation level, and then the spin correlation can be investigated on the phase diagram, and specially, along the freeze-out line to compare with the experimental data of Beam Energy Scan, and we have chance to investigate how crossover and Critical End Point influence the spin correlation.

This paper is organized as follows. Section~\ref{sec:definition} defines the observable and reduces it to the quark level. Section~\ref{sec:yukawatheory} derives the spin weights and the orbital barrier in a general Yukawa theory, in which the scalar and pseudo-scalar mesons are represented as point particles, and Sec.~\ref{sec:qm} identifies the underlying total-spin states. In Sec.~\ref{sec:njl} we construct the three-flavor NJL description and the RPA spectral functions, replacing the point-particle mesons by these more realistic ones. Section~\ref{sec:results} presents the numerical results, and Sec.~\ref{sec:conclusions} summarizes our findings.

\section{Spin correlation definition}
\label{sec:definition}

Microscopically, the spin correlation of an $s\bar{s}$ pair along a common spin-analysis axis $\hat{\mathbf{a}}$ is characterized by the two-spin expectation value:
\begin{equation}
   P_{s\bar{s}}(\hat{\mathbf{a}})=\frac{
\operatorname{Tr}\left[\rho(\mathbf{p},\mathbf{k})\biggl((\boldsymbol{\sigma}_{s}\cdot\hat{\mathbf{a}})\otimes(\boldsymbol{\sigma}_{\bar{s}}\cdot\hat{\mathbf{a}})\biggr)\right]}{\operatorname{Tr}\rho(\mathbf{p},\mathbf{k})},
\label{eq:Pdef0}
\end{equation}
which measures the difference between the probabilities of parallel and antiparallel spin configurations along this axis. Here $\boldsymbol{\sigma}_{s/\bar{s}}$ are the Pauli matrices acting in the spin space of the quark and of the antiquark, the trace is taken over that two-particle spin space, and $\rho(\mathbf{p},\mathbf{k})$ is the two-particle spin density matrix for a quark--antiquark pair with momenta $\mathbf{p}$ and $\mathbf{k}$:
\begin{equation}
    \rho_{sr,s'r'}(\mathbf{p},\mathbf{k})=\left\langle
a^{\dagger}_{s'}(\mathbf{p})\,
b^{\dagger}_{r'}(\mathbf{k})\,
b_{r}(\mathbf{k})\,
a_{s}(\mathbf{p})
\right\rangle,
\end{equation}
where $a_s(\mathbf{p})$ and $b_r(\mathbf{k})$ annihilate a strange quark and a strange antiquark, respectively, and $s,r,s',r'\in\{\uparrow,\downarrow\}$ label their spin states along axis $\hat{\mathbf{a}}$. It is therefore natural to expand the two-particle density matrix in basis
$\{\mathbb{I}_{2\times2},\sigma_{i}\}\otimes\{\mathbb{I}_{2\times2},\sigma_{j}\}$:
\begin{align}
\rho_{sr,s'r'}&=\frac{1}{4}\mathcal{D}\,\delta_{ss'}\delta_{rr'}
+\frac{1}{4}\sum_{i,j=1}^{3}\mathcal{N}_{ij}\,(\sigma_{i})_{ss'}(\sigma_{j})_{rr'}\nonumber\\
&\quad+\frac{1}{4}\sum_{i=1}^{3}\Big[\mathcal{S}_{i}\,(\sigma_{i})_{ss'}\delta_{rr'}
+\bar{\mathcal{S}}_{i}\,\delta_{ss'}(\sigma_{i})_{rr'}\Big],
\label{eq:rhoexpansion}
\end{align}
in which the first term is the spin-independent, the second is the two-spin correlation term, and the last collects the two single-spin polarization, which do not contribute to $P_{s\bar{s}}$. Using the trace orthogonality $\Tr(\sigma_{i}\sigma_{j})=2\delta_{ij}$, the coefficients of non-correlated term can be extracted by:
\begin{align}
    \mathcal{D}=\Tr\rho=\sum_{s,r}\rho_{sr,sr},
    \label{eq:D}
\end{align}
which is exactly the total pair yield, and the spin-correlated term:
\begin{align}
    \!\!\!\!\!\!\!\mathcal{N}_{ij}&=\Tr\big[\rho\,(\sigma_{i}\otimes\sigma_{j})\big]=\sum_{s,s'}\sum_{r,r'}\rho_{sr,s'r'}(\sigma_{i})_{s's}(\sigma_{j})_{r'r},
\label{eq:Nij}
\end{align}
then substitute Eq.~\eqref{eq:rhoexpansion} into Eq.~\eqref{eq:Pdef0} and use $\Tr\sigma_{i}=0$,  the observable collapses to a single contraction of the correlation tensor,
\begin{align}
    P_{s\bar{s}}(\hat{\mathbf{a}})=\sum_{i,j=1}^{3}\hat a_{i}\hat a_{j}\,\mathcal{C}_{ij},\quad \mathcal{C}_{ij}\equiv\frac{\mathcal{N}_{ij}}{\mathcal{D}}.
\label{eq:Pchain}
\end{align}
Eq.~\eqref{eq:Pchain} is exact and involves no kinematic assumption about the decay. Since $\mathcal{C}_{ij}$ is a ratio, any overall normalization of the pair yield cancels. Any spin-independent contribution to $\rho$, such as the uncorrelated thermal pairs that form the background, has $\mathcal{N}_{ij}=0$ and enters through $\mathcal{D}$ alone.

What the experiment measures is the space-averaged spin correlation, while Eq.~\eqref{eq:Pchain} still depends on the orientation of the pair relative to the fixed laboratory axis $\hat{\mathbf{a}}$.  Notice that
$\sum_{i}\hat a_{i}^{2}=\sum_{j}\hat a_{j}^{2}=1$ and the standard rotation average
$\langle \hat a_{i}\hat a_{j}\rangle_{\rm iso}=\tfrac13\delta_{ij}$, then finally we obtain the spin correlation of strange quark and anti-quark:
\begin{equation}
P_{s\bar{s}}=\frac{1}{3}\sum_{i=1}^{3}\mathcal{C}_{ii}
=\frac{1}{3}\operatorname{Tr}\mathcal{C}.
\label{eq:Ptrace}
\end{equation}
Now the core task is to determine $\mathcal{N}_{ij}$ and $\mathcal{D}$ in given theory.

\section{Spin Correlation In Yukawa Theory}
\label{sec:yukawatheory}
\subsection{Basic derivation}
We first investigate the spin correlation in a general theory with a Yukawa-like interaction, in which the meson is represented as a point particle and the interaction between the quark and the meson takes the form
\begin{align}
     \mathcal{L}_{\rm int}(x) = -g_{\Phi ss} \Phi(x) \bar{s}(x)\Gamma s(x),
\label{eq:int}
\end{align}
$\Phi$ is the meson field which could be scalar meson $\Gamma=\mathbf{1}$ or pseudo-scalar meson $\Gamma=i\gamma^5$. Now consider meson melting into a quark-antiquark pair at finite temperature:
\begin{align}
    \Phi (k) \rightarrow s(p,r_1) + \bar{s}(k,s_2),
\end{align}
the joint transition probability tensor at tree level in the spin space of the final state is the outer product of the decay amplitudes,
\begin{align}
\rho_{sr,s'r'}& =  \mathcal{M}_{r_1, s_2} \mathcal{M}^*_{r_1', s_2'} \nonumber\\
&= g^2 \Big[ \bar{u}^{r_1}(p) \Gamma v^{s_2}(k) \Big] \Big[ \bar{v}^{s_2'}(k) \Gamma u^{r_1'}(p) \Big].
\label{eq:Tdef}
\end{align}
with the relation in Eq.~\eqref{eq:D} and \eqref{eq:Nij}, and using the covariant completeness relations
\begin{align}
\sum_{r_1, r_1'} u^{r_1'}(p) (\sigma_i)_{r_1' r_1} \bar{u}^{r_1}(p) &= (\rlap/{p} + m) \gamma_5 \rlap/{\hat{e}}_{(i)}(p),\nonumber \\
\sum_{s_2, s_2'} v^{s_2}(k) (\sigma_j)_{s_2' s_2} \bar{v}^{s_2'}(k) &= (\rlap/{k} - m) \gamma_5 \rlap/{\hat{n}}_{(j)}(k),
\end{align}
together with the standard spin sums $\sum_r u^r\bar{u}^r=\rlap/p+m$ and
$\sum_s v^s\bar{v}^s=\rlap/k-m$, then we obtain:
\begin{align}
\mathcal{D} &= g^2 \operatorname{Tr} \Big[ (\rlap/{p} + m) \Gamma (\rlap/{k} - m) \Gamma \Big], \\
\mathcal{N}_{ij} &= g^2 \operatorname{Tr} \Big[ (\rlap/{p} + m) \gamma_5 \rlap/{\hat{e}}_{(i)} \, \Gamma \, (\rlap/{k} - m) \gamma_5 \rlap/{\hat{n}}_{(j)} \, \Gamma \Big],
\label{eq:tracereps}
\end{align}
with $\hat{e}_{(i)}$, $\hat{n}_{(j)}$ orthonormal spacelike triads orthogonal to $p$ and $k$, $\hat e_{(i)}\cdot\hat e_{(j)}=\hat n_{(i)}\cdot\hat n_{(j)}=-\delta_{ij}$. Performing the Dirac traces of Eq.~\eqref{eq:tracereps} for $\Gamma=\mathbf{1}$ and $\Gamma=i\gamma^{5}$, contracting the triads, and summing over the final-state spins, one obtains for each channel both the trace of the numerator tensor and the unpolarized rate, $\mathcal{N}_{\sigma/\pi}\equiv\Tr\,\mathcal{N}_{ij}^{\sigma/\pi}$. Here and below the subscripts $\sigma$ and $\pi$ label the scalar and pseudo-scalar channels rather than the physical $\sigma$ and $\pi$ mesons, since our focus is on the spin correlations of strange quarks:
\begin{align}
    \mathcal{N}_{\sigma}
      &= 4A_{\sigma}g_{\sigma}^2\,\mathcal{P}_{-}\,\mathcal{B},\quad\mathcal{D}_{\sigma}=4A_{\sigma}g_{\sigma}^2\,\mathcal{P}_{-}\,\mathcal{B},\nonumber\\
    \mathcal{N}_{\pi}
       &= -12A_{\pi}g_{\pi}^2\,\mathcal{P}_{+}\,\mathcal{B},\quad
\mathcal{D}_{\pi}=4A_{\pi}g_{\pi}^2\,\mathcal{P}_{+}\,\mathcal{B},
    \label{eq:ratemeson}
\end{align}
where $\mathcal{P}_{\pm}\equiv p\cdot k\pm m^2$, $A_{M}(\omega,\mathbf{k})=\rho_{M}(\omega,\mathbf{k})\,n_B(\omega)$, with the spectral function $\rho_{M}$ and Bose distribution $n_B(\omega)=1/(e^{\omega/T}-1)$, and the final-state Pauli factor $\mathcal{B}\equiv\big[1-f^{-}(E_{1})\big]\big[1-f^{+}(E_{2})\big]$, which is the probability that both final-state states are unoccupied, and the Fermi-Dirac distribution $f^{\pm}(E)=1/(e^{(E\pm\mu_{q})/T}+1)$.  The Dirac structure of the two vertices is what fixes the relative weight, and the traces of the single-channel correlation tensors are pure numbers,
\begin{equation}
\operatorname{Tr}\mathcal{C}_{\sigma}
=\frac{\mathcal{N}_{\sigma}}{\mathcal{D}_{\sigma}}=+1,
\qquad
\operatorname{Tr}\mathcal{C}_{\pi}
=\frac{\mathcal{N}_{\pi}}{\mathcal{D}_{\pi}}=-3 .
\label{eq:TrC}
\end{equation}
The two signs are opposite, which makes the two channels compete in spin
correlation.

On the other hand, in a real heavy ion collision the scalar and pseudo-scalar mesons both contribute to the spin correlation simultaneously, while spin-unrelated quark pairs can also be produced thermally through other channels as well as in hard processes. Since this is quite complicated and goes beyond the scope of this article, for simplicity we phenomenologically introduce a spin-blind background $B_{0}$: uncorrelated thermal $q\bar{q}$ pairs in the spin--colour channels that the two mesonic channels do not already account for. The spin correlation then reads
\begin{align}
    P_{s\bar{s}}&=\frac{1}{3}\,
                \frac{\mathcal{N}_{\sigma}+\mathcal{N}_{\pi}}
                     {\mathcal{D}_{\sigma}+\mathcal{D}_{\pi}+B_0 f^{-}(E_1)f^{+}(E_2)}\nonumber\\
&=\frac{\rho_{\sigma}\,\mathcal{P}_{-}-3\rho_{\pi}\,\mathcal{P}_{+}}{3\rho_{\pi}\,\mathcal{P}_{+}+3\rho_{\sigma}\,\mathcal{P}_{-}+3B_0\big/4(1+n_{B})},
    \label{eq:Pmaster}
\end{align}
where we have used detailed-balance identity $n_{B}(\omega)\,\mathcal{B}\;=\;\big[1+n_{B}(\omega)\big]\,f^{-}(E_{1})f^{+}(E_{2})$.
In Eq.~\eqref{eq:Pmaster}, we have omitted the coupling constant $g^2_{\sigma/\pi}$, because it is absorbed into the corresponding spectral function $\rho_{\sigma/\pi}$ in the NJL model, as shown in Sec.~\ref{sec:njl}. 
On the other hand, for a color-singlet pair the degeneracy is $(2s_{1}+1)(2s_{2}+1)N_{c}=12$; subtracting the scalar and pseudo-scalar units that are treated dynamically leaves a residual $B_{0}=12-2=10$, and it dilutes $P$ without changing its sign. 


And it's clear that scalar and pseudo-scalar meson contribute totally opposite to the spin correlation from Eq.~\eqref{eq:Pmaster}, and two opposite limits follow immediately:
At $\theta=0$ the pair is collinear, $\mathcal{P}_{-}=2|\mathbf{p}|^{2}\sin^{2}(\theta/2)$ vanishes, and the $\sigma$ channel drops out entirely, then
    \begin{equation}
        P_{s\bar{s}} =-\frac{3\rho_{\pi}\,\mathcal{P}_{+}}{3\rho_{\pi}\,\mathcal{P}_{+}+3B_0\big/4(1+n_{B})},
        \label{eq:Pcollinear}
    \end{equation}
the pseudo-scalar channel alone survives.  At $\theta=\pi$ the pair is back-to-back, $\mathbf{q}=\mathbf{0}$, and $\mathcal{P}_{-}=2|\mathbf{p}|^2$. Now both channels contribute,
        \begin{equation}
            P_{s\bar{s}} =\frac{\rho_{\sigma}|\mathbf{p}|^2-3\rho_{\pi}E_{\mathbf{p}}^2}{3\rho_{\pi}E_{\mathbf{p}}^2+3\rho_{\sigma}|\mathbf{p}|^2+3B_0\big/8(1+n_{B})},
            \label{eq:Pbacktoback}
        \end{equation}
with $E_{\mathbf{p}}^{2}=|\mathbf{p}|^2+m^{2}$. This is the configuration in which the scalar channel competes most strongly, and most so at small $|\mathbf{p}|$. In the degenerate limit $\mathbf{p}=0$ the scalar channel again switches off, and the background alone sets a floor on $|P|$:
    \begin{align}
         P_{s\bar{s}}(\mathbf{p}{=}0) =-\frac{1}{1+B_0\big/8\rho_{\pi}m^2(1+n_{B})},
         \label{eq:Pzerop}
     \end{align}
which equals $-1$ at $B_0=0$ and rises toward zero as $B_0$ increases. The background therefore weakens spin correlation signals and sets the low-momentum floor of $|P_{s\bar{s}}|$.

\subsection{Quantum-Mechanical Origin of Spin Correlation}
\label{sec:qm}

Let us now consider, from a quantum-mechanical perspective, how the above spin correlations arise. Conservation of $\mathbf{J}=\mathbf{L}+\mathbf{S}$ in the decay of a $J=0$ meson forces $L=S$, while the pair parity is $(-1)^{L+1}$; the two vertices of Eq.~\eqref{eq:int} therefore select
\begin{align}
\Gamma=\mathbf{1}\ (0^{+}):\quad
  (-1)^{L+1}=+1\ &\Rightarrow\ L=S=1\;(^{3}P_{0}),\nonumber\\
\Gamma=i\gamma^{5}\ (0^{-}):\quad
  (-1)^{L+1}=-1\ &\Rightarrow\ L=S=0\;(^{1}S_{0}).\nonumber
\end{align}
The scalar vertex produces the pair in a spin triplet, while the pseudo-scalar vertex in a singlet. Writing the two spin states along momentum $\hat{\mathbf{p}}$ as
$\uparrow,\downarrow$, the vertices give
\begin{equation}
\!\!\!\!\!|^{3}P_{0}\rangle=\frac{1}{\sqrt2}
  \big(|\!\uparrow\downarrow\rangle+|\!\downarrow\uparrow\rangle\big),
\,
|^{1}S_{0}\rangle=\frac{1}{\sqrt2}
  \big(|\!\uparrow\downarrow\rangle-|\!\downarrow\uparrow\rangle\big).
\label{eq:states}
\end{equation}
The spin correlation measured along a common axis $\hat{\mathbf a}$ is defined as
\begin{align}
    P(\hat{\mathbf a};\hat{\mathbf p})=\left\langle
(\boldsymbol\sigma_1\cdot\hat{\mathbf a})
(\boldsymbol\sigma_2\cdot\hat{\mathbf a})
\right\rangle.
\end{align}
Averaging over the unmeasured pair orientation, as in
Eq.~\eqref{eq:Ptrace}, the three spatial directions contribute equally and
\begin{equation}
\overline{\langle (\boldsymbol{\sigma}_{1}\!\cdot\!\hat{\mathbf{a}})
(\boldsymbol{\sigma}_{2}\!\cdot\!\hat{\mathbf{a}})\rangle_{\hat{\mathbf{p}}}}^{\ \rm iso}
=\frac{1}{3}\,\boldsymbol{\sigma}_{1}\!\cdot\!\boldsymbol{\sigma}_{2}
=\frac{2S(S+1)-3}{3},
\label{eq:SdotS}
\end{equation}
using $\boldsymbol{\sigma}_{1}\!\cdot\!\boldsymbol{\sigma}_{2}
=2\mathbf{S}^{2}-3$ for two spin-$\tfrac12$ particles. For the two states of
Eq.~\eqref{eq:states},
\begin{equation}
P_{s\bar{s}}=
\begin{cases}
+\dfrac{1}{3}, & \text{scalar }(^{3}P_{0},\; S=1),\\[8pt]
-1, & \text{pseudo-scalar }(^{1}S_{0},\; S=0).
\end{cases}
\label{eq:PS}
\end{equation}

The two states respond to the analysis axis $\hat{\mathbf{a}}$ in qualitatively different ways,
\begin{equation}
P_{\sigma}(\hat{\mathbf{a}})
=1-2(\hat{\mathbf{a}}\!\cdot\!\hat{\mathbf{p}})^{2},
\qquad
P_{\pi}(\hat{\mathbf{a}})=-1 .
\label{eq:peraxis}
\end{equation}
The pseudo-scalar channel is isotropic, the singlet being rotationally invariant: it gives $P=-1$ along every axis. The scalar channel is quadrupolar: antiferromagnetic along the pair axis ($P=-1$ at $\hat{\mathbf{a}}\parallel\hat{\mathbf{p}}$), ferromagnetic transverse to it ($P=+1$), vanishing at $(\hat{\mathbf{a}}\!\cdot\!\hat{\mathbf{p}})^{2}=\tfrac12$ and averaging to the $+\tfrac13$ of Eq.~\eqref{eq:PS}. The two channels thus differ not only in sign but in shape: the pseudo-scalar contribution is constant in angle, the
scalar one a quadrupole oriented along the pair.

\subsection{Competition of Scalar and Pseudo-Scalar Channel}
The actual computation now reduces to a competition between two spectral functions evaluated along the pair kinematics, which requires three ingredients. 

First, the scalar channel is suppressed relative to the pseudo-scalar one by its $1{:}3$ numerator weight. Second, it is suppressed further by the orbital barrier. In the pair rest frame the combinations of Eq.~\eqref{eq:ratemeson} are simply
\begin{equation}
\mathcal{P}_{-}=2|\mathbf{p}|^{2},\quad\mathcal{P}_{+}=2E_{\mathbf{p}}^{2},
\label{eq:orbital}
\end{equation}
The origin of the orbital barrier is then transparent. A pair with relative orbital angular momentum $L$ feels the centrifugal potential $\propto L(L+1)/r^{2}$, which is repulsive and grows without bound as the separation $r\to0$. As $|\mathbf{p}|\to0$ the barrier becomes impenetrable, $\mathcal{P}_{-}$ tends to vanish, and so does the contribution from the scalar channel, while $\mathcal{P}_{+}$ carries no barrier and serves as the only contribution. The angular dependence of the observable is thus the signature of the total spin the vertex selects, filtered by the orbital barrier it has to overcome. Third, the two channels enter with their spectral weights $\rho_{\sigma}$ and $\rho_{\pi}$, a ratio which turns out to be of order unity and even larger than unity over much of the sampled window (as shown in Fig.~\ref{fig:spectral} below).

In the medium frame the ratio of the two factors is
\begin{equation}
\frac{\mathcal{P}_{-}}{\mathcal{P}_{+}}=1-\frac{4m^{2}}{s},
\end{equation}
Combining the three factors above, we write
\begin{equation}
P_{s\bar{s}}=-\frac{1-R}{1+3R},
\qquad
R=\frac{1}{3}\times \frac{\mathcal{P}_{-}}{\mathcal{P}_{+}}\times\frac{g^2_{\sigma}\rho_{\sigma}}{g^2_{\pi}\rho_{\pi}},
\end{equation}
where the $1/3$ is the spin-weight ratio, the second factor is the orbital barrier of Eq.~\eqref{eq:orbital}, and the third collects the spectral weights. The numerical evaluation of Sec.~\ref{sec:results} gives $R<1$ in the kinematically accessible region, so that the pseudo-scalar channel dominates and $P_{s\bar{s}}$ remains negative throughout.


\section{Three-flavor NJL model and the RPA meson propagator}
\label{sec:njl}

We now embed the standalone Yukawa framework into the three-flavor NJL model~\cite{Nambu:1961tp,Klevansky:1992qe,Hatsuda:1994pi} with the Lagrangian
\begin{align}
    \mathcal{L}&=\bar\psi(i\slashed\partial+\mu\gamma^{0}-\hat m_{q})\psi\nonumber\\
&+G\sum_{a=0}^{8}\big[(\bar\psi\lambda^{a}\psi)^{2}+(\bar\psi i\gamma^{5}\lambda^{a}\psi)^{2}\big]\nonumber\\
&-K\Big(\det_{i,j}\big[\bar\psi_i(1+\gamma^{5})\psi_j\big]
      +\det_{i,j}\big[\bar\psi_i(1-\gamma^{5})\psi_j\big]\Big),
\label{eq:njl-lagrangian}
\end{align}
where $\psi=(u,d,s)^{T}$ is the quark field, $\hat m_{q}=\mathrm{diag}(m_{u},m_{d},m_{s})$ is the current mass, $G$ is the $U(3)_{L}\times U(3)_{R}$ symmetric scalar coupling, and $K$ is the $U_{A}(1)$-breaking 't~Hooft determinant coupling that induces flavor mixing among $u,d,s$ quarks. In the mean-field approximation, the constituent quark masses are determined self-consistently from the gap equations
\begin{equation}
m_{f}=m_{f,0}-4G\,\sigma_{f}+2K\,\sigma_{f+1}\,\sigma_{f+2},
\label{eq:gap}
\end{equation}
where $\sigma_{f}=\langle\bar f f\rangle$ is the quark condensate and can be expressed by the gap equation:
\begin{equation}
    \sigma_{f}=-\frac{N_{c}}{2\pi^{2}}\!\int\frac{\mathbf{p}^{2}d\mathbf{p}\,m_{f}}{E_{\mathbf{p},f}}
    \big[1-f^{+}(E_{\mathbf{p},f})-f^{-}(E_{\mathbf{p},f})\big],
\end{equation}
with cyclic flavor indices $(u\to d\to s\to u)$ and  $E_{\mathbf{p},f}=\sqrt{\mathbf{p}^{2}+m_{f}^{2}}$.

Assuming isospin symmetry ($m_{u}=m_{d}$, $\sigma_{u}=\sigma_{d}\equiv\sigma_{l}$), we transform to the light--strange ($l$--$s$) basis with scalar currents $J_{l}=(\bar u u+\bar d d)/\sqrt{2}$ and $J_{s}=\bar s s$. The scalar-channel interaction then reads
\begin{align}
\mathcal{L}_{\rm int}^{\rm scalar}=G_{l}\,J_{l}^{2}+G_{s}\,J_{s}^{2}+2G_{ls}\,J_{l}J_{s},
\label{eq:Lscalar}
\end{align}
where the effective couplings are dynamically dressed by the condensates $G_{l}=2G-K\sigma_{s},\quad G_{s}=2G,\quad G_{ls}=-\sqrt{2}\,K\,\sigma_{l}$. The off-diagonal $G_{ls}$ explicitly shows that the 't~Hooft term bridges the strange and light sectors. As chiral symmetry is restored, $G_{ls}\to0$ and the light--strange mixing is switched off.

The RPA resummation of the scalar bubble chain generates a matrix propagator in the $(l,s)$ basis,
\begin{align}
\mathbb{D}(q)=2\mathbb{G}\big[\mathbbm{1}-2\,\mathbf{\Pi}(q)\,\mathbb{G}\big]^{-1},\,\mathbb{G}=\begin{pmatrix}G_{l}&G_{ls}\\G_{ls}&G_{s}\end{pmatrix},
\label{eq:Dmatrix}
\end{align}
where $ \mathbf{\Pi}=\mathrm{diag}(\Pi_{l},\Pi_{s})$ and $\Pi_{f}(q)$ is the one-loop scalar polarization of flavor $f$. Projecting onto the $s\bar{s}$ channel yields 
\begin{equation}
    \mathcal{D}_{ss}(q^{2})=\frac{N_{s}(q^{2})}{\Delta(q^{2})}
\end{equation}
with 
\begin{align}
     N_{s}(q^{2})&=2\big[G_{s}(1-2G_{l}\Pi_{l})+2G_{ls}^{2}\,\Pi_{l}\big],\\
      \Delta(q^{2})&=1-2G_{l}\Pi_{l}-2G_{s}\Pi_{s}+4(G_{l}G_{s}-G_{ls}^{2})\Pi_{l}\Pi_{s},\nonumber\\
\end{align}
and the zeros of $\Delta(q^{2}{=}0,i\nu_{n}{=}0; T,\mu_{B})$ give the in-medium $\sigma$ screening mass.

In direct analogy, the iso-singlet pseudo-scalar sector is built from the currents
$J_{l}^{P}=(\bar u i\gamma^{5} u+\bar d i\gamma^{5} d)/\sqrt{2}$ and
$J_{s}^{P}=\bar s i\gamma^{5} s$ with the same structure, except that the parity
property of the 't~Hooft determinant flips the sign of the
condensate-dressed anomaly, $G_{l}^{P}=2G+K\sigma_{s}$, $G_{s}^{P}=2G$,
$G_{ls}^{P}=+\sqrt{2}K\sigma_{l}$. The off-diagonal term governs the in-medium
mixing between the light and strange pseudo-scalar currents that form the
physical $\eta$ and $\eta'$. The RPA resummation and the pole residues follow the same pattern, $\mathbb{D}^{P}=2\mathbb{G}^{P}[\mathbbm{1}-2\mathbf{\Pi}^{P}
\mathbb{G}^{P}]^{-1}$ and  $\mathcal{D}_{ss}^{P}=N_{s}^{P}/\Delta^{P}$. More details of above calculation can be found in Ref.\cite{Hansen:2006ee,Rehberg:1995kh}

However, to reach the spin correlation defined above, what we really need the spectral function at given temperature and baryon chemical potential, as shown in previous section. Writing a general channel as $X\in\{\text{scalar}, \text{pseudo}\}$ (superscript $S$ for the scalar sector while $P$ for the pseudo-scalar sector), the flavor-diagonal polarizations are complex across the physical
continuum, $\Pi_{f}^{X}=\mathrm{Re}\,\Pi_{f}^{X}
+i\,\mathrm{Im}\,\Pi_{f}^{X}$, then the spectral function projected onto the
$s\bar{s}$ source channel is
\begin{align}
\rho_{ss}^{X}(q) = 2\,\frac{\mathrm{Im}\,N_{s}^{X}\,\mathrm{Re}\,\Delta^{X} - \mathrm{Re}\,N_{s}^{X}\,\mathrm{Im}\,\Delta^{X}}{\big[\mathrm{Re}\,\Delta^{X}\big]^{2} + \big[\mathrm{Im}\,\Delta^{X}\big]^{2}},
\label{eq:spectral_full}
\end{align}
thus $\rho_{ss}^{S}$ automatically encompasses the $\sigma$ and $f_{0}$ modes together with their multi-quark threshold continua, while $\rho_{ss}^{P}$ captures the coupled isosinglet $\eta$--$\eta'$ excitations. 

The building blocks entering the RPA spectral functions are the one-loop
polarization functions $\Pi_{f}^{X}(q)$ for a quark of flavor
$f\in\{l,s\}$ with constituent mass $m_f$ and effective chemical potential
$\mu_f$. For $q^{\mu}=(\omega,\mathbf{q})$ with $q^{2}=\omega^{2}
-\mathbf{q}^{2}$, they are evaluated in the imaginary-time Matsubara formalism
as 
\begin{align}
   \!\!\!\! \Pi_{f}^{X}(i\nu_n,\mathbf{q})=-T\sum_{n}\int\frac{\mathrm{d}^{3}\mathbf{k}}{(2\pi)^{3}} \mathrm{Tr}[\Gamma_{X}
S_{f}(k+q)\Gamma_{X}S_{f}(k)]\nonumber\\
\end{align}
where $\nu_n=2n\pi T$ is the bosonic Matsubara frequency, $S_f(k)=(\rlap/ k-m_f+\mu_f\gamma^{0})^{-1}$ is the in-medium quark
propagator, and the vertices are $\Gamma_{\rm scalar}=\mathbb{I}$ and $\Gamma_{\rm pseudo}=i\gamma^{5}$. Performing the Dirac traces and summing over the Matsubara frequencies, the analytic continuation $i\nu_n\to\omega+i\epsilon$ casts both
channels into a unified structure,
\begin{align}
\Pi_{f}^S(q)   &= I_0^f + (q^2 - 4m_f^2)\, I_f(\omega, \mathbf{q}), \nonumber \\
\Pi_{f}^P(q) &= I_0^f + q^2\, I_f(\omega, \mathbf{q}), 
\label{eq:Pi_compact}
\end{align}
where the dynamic kernel reads
\begin{align}
    I_f = N_c \sum_{s,s'} \int \frac{\mathrm{d}^3\mathbf{k}}{(2\pi)^3}\,\frac{1}{4E_{\mathbf{k}}E_{\mathbf{p}}} \frac{1 - f_{\mathbf{k}}^{s} - f_{\mathbf{p}}^{s'}}{\omega - s E_{\mathbf{k}} - s' E_{\mathbf{p}}} ,
\label{eq:dynamical_kernel_compact}
\end{align}
with $E_{\mathbf{p}} = E_{\mathbf{k}+\mathbf{q}}$, $f_k^{s} \equiv f(E_\mathbf{k} - s\mu_f) $. And now the imaginary parts :
\begin{align}
\!\!\!\!\mathrm{Im}\,\Pi_{f}^S(\omega, \mathbf{q})   &= \frac{N_c\,(q^2 - 4m_f^2)}{8\pi |\mathbf{q}|}\,\mathcal{F}(\omega,q)\,\Theta(q^2 - 4m_f^2), \nonumber \\
\!\!\!\!\mathrm{Im}\,\Pi_{f}^P(\omega, \mathbf{q}) &= \frac{N_c\,q^2}{8\pi |\mathbf{q}|}\,\mathcal{F}(\omega,q)\,\Theta(q^2 - 4m_f^2), \label{eq:ImPi}
\end{align}
where
\begin{equation}
\mathcal{F}(\omega,q)\equiv\int_{E_-}^{E_+}\! \mathrm{d}E_{\mathbf{k}}\,\big[ 1 - f_F^+(E_{\mathbf{k}}) - f_F^-(\omega - E_{\mathbf{k}}) \big]
\label{eq:F_def}
\end{equation}
is the thermal phase-space integral and the boundaries are kinematically bounded by
\begin{equation}
E_\pm = \frac{\omega}{2} \pm \frac{|\mathbf{q}|}{2}\sqrt{1 - \frac{4m_f^2}{q^2}}.
\end{equation}

\section{Numerical results}
\label{sec:results}

We now evaluate the framework above with the three-flavor NJL parameters of Hatsuda and Kunihiro~\cite{Hatsuda:1994pi}: the current quark mass $m_{u,0}=m_{d,0}=5.5$~MeV, $m_{s,0}=140.7$~MeV, coupling constants $G\Lambda^{2}=1.835$, $K\Lambda^{5}=12.36$, as well as the three-momentum cutoff  $\Lambda=602.3$~MeV, which are fitted to reproduce $f_{\pi}=92.4$~MeV, $m_{\pi}=135.0$~MeV, $m_{K}=497.7$~MeV and $m_{\eta'}=957.8$~MeV in vacuum. And within these parameters,  it is found the chiral crossover temperature $T_{c}(0)=172.9$~MeV at vanishing chemical potential,  which is located by the peak of the light-quark chiral susceptibility $\chi_{l}=-\partial M_{l}/\partial T$,  and the critical end point locates at $T_{\rm CEP}=68.2$~MeV and $\mu_{B,\rm CEP}=948$~MeV. 

\begin{figure}[htbp]
\centering
\includegraphics[width=\columnwidth]{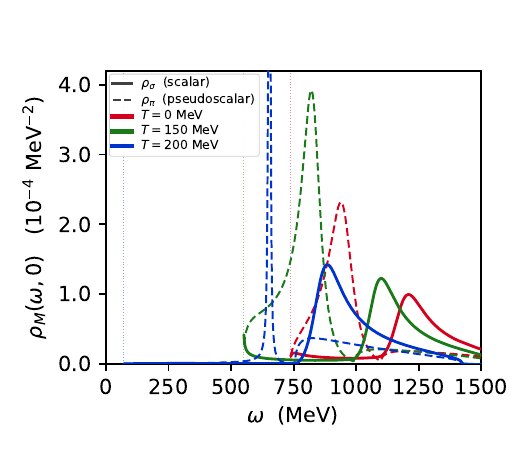}
\caption{The scalar ($\rho_{\sigma}$, solid) and pseudo-scalar ($\rho_{\pi}$,
dashed) spectral functions of Eq.~\eqref{eq:spectral_full} at $\mathbf{q}=0$ and $\mu_{B}=0$, for $T=0$ (red), $150$~MeV (green) and $200$~MeV (blue). The vertical dotted lines mark the two-quark thresholds $2M_{l}$ of each temperature. Curves are drawn from
$2M_{l}$ upwards.}
\label{fig:spectral}
\end{figure}

We also show the spectral functions in Fig.~\ref{fig:spectral} for three temperatures and at vanishing chemical potential, taking $\mathbf{q}=0$, which is the configuration entering the back-to-back observable. The curves are drawn only above the two-quark threshold $2M_{l}$ of the light sector; below it both spectral functions vanish identically, as the step function in Eq.~\eqref{eq:ImPi} requires.
Three features of Fig.~\ref{fig:spectral} are relevant for the observable.
First, both peaks move downwards as the temperature rises, which is the expected behavior
of chiral restoration that the in-medium masses track the melting condensate. Second, the pseudo-scalar mode lies below the scalar one at every temperature, so that the ratio $\rho_{\sigma}/\rho_{\pi}$ is not uniformly below unity at the kinematically accessible energies, $\omega\ge2M_{s}$, where the $s\bar{s}$ pair can actually be created. The spectral factor therefore does not act as a suppression of the scalar channel. 

A further point concerns the poles themselves. Because the observable samples
$\omega\ge2M_{s}$, whereas the $\eta$ and $f_0$ peaks sit at
$\omega\simeq938$ and $1210$~MeV at $T=0$ and $2M_{s}=1099$~MeV, the peaks lie
at or below the lower edge of the sampled window. The fraction $P_{s\bar{s}}$
is thus sensitive not to the resonance maxima but to the high-$\omega$ tails of
both spectral functions. This is precisely why the observable maps the chiral
crossover rather than the critical end point: whatever structure the meson
poles develop near the CEP, it is filtered out by the $2M_{s}$ threshold before
it can reach $P_{s\bar{s}}$.

\subsection{The phase-space measure}
\label{subsec:measure}
Before showing the numerical result of spin correlation,  we have to make it clear how to calculate the averaged spin correlation over phase space. Notice that the pair is produced through a meson mode of four-momentum $k^{\mu}$, which four-momentum conservation at the vertex identifies with the total momentum of the pair,
\begin{equation}
k^{\mu}=Q^{\mu}\equiv p_{1}^{\mu}+p_{2}^{\mu}=(\omega,\mathbf{q}),
\end{equation}
with $\omega=E_{1}+E_{2}$ and $\mathbf{q}=\mathbf{p}+\mathbf{k}$. The pair is thus specified by $s\equiv Q^{2}=\omega^{2}-\mathbf{q}^{2}$, $\omega=\sqrt{s+\mathbf{q}^{2}}$, where $s$ is the invariant mass squared of the pair and sets the threshold, while $\omega$ is its energy in the medium frame and fixes the thermal occupation. Integrating the quark momenta in the pair rest frame ($E_{1}=E_{2}=\sqrt{s}/2$) leaves the two-body phase space:
\begin{align}
\Phi_{2}&=\int\!\frac{\mathrm{d}^{3}p_{1}}{(2\pi)^{3}2E_{1}}
                \frac{\mathrm{d}^{3}p_{2}}{(2\pi)^{3}2E_{2}}\,
                (2\pi)^{4}\delta^{4}(Q-p_{1}-p_{2})\nonumber\\[2pt]
 &=\frac{1}{8\pi}\beta(s),
\label{eq:Phi2}
\end{align}
with $\beta(s)=\sqrt{1-4m^{2}/s}$. 

The spectral representation of the meson propagator supplies the mode occupation $n_{B}(\omega)$ and its off-shell weight $\rho_{M}(\omega,\mathbf{q})$, and notice $\mathrm{d}\omega=\mathrm{d}s/(2\omega)$, thus the pair yield becomes
\begin{equation}
\mathrm{d}N=\frac{1}{32\pi^{4}}\,
n_{B}(\omega)\,\beta(s)\,\rho_{M}(\omega,\mathbf{q})\,
\frac{\mathbf{q}^{2}}{2\omega}\,
\mathrm{d}s\,\mathrm{d}|\mathbf{q}|.
\label{eq:measure2}
\end{equation}
The observable is the ratio of the numerator and denominator of Eq.~\eqref{eq:Pmaster} integrated over this yield,
\begin{equation}
\bar{P}_{s\bar{s}}
=\frac{1}{3}\frac{\displaystyle\int \mathrm{d}\mu\;\mathcal{N}}{\displaystyle\int \mathrm{d}\mu\;\mathcal{D}},
\quad
\mathrm{d}\mu\equiv
n_{B}(\omega)\,\mathcal{B}\,\beta(s)\,
\frac{\mathbf{q}^{2}}{2\omega}\,
\mathrm{d}s\,\mathrm{d}|\mathbf{q}| ,
\label{eq:P2d}
\end{equation}
where
\begin{align}
    \mathcal{N}&=\rho_{\sigma}\mathcal{P}_{-}-3\rho_{\pi}\mathcal{P}_{+},\nonumber\\
\mathcal{D}&=\rho_{\pi}\mathcal{P}_{+}+\rho_{\sigma}\mathcal{P}_{-}
+\frac{B_0}{4(1+n_{B})} .
\label{eq:P2dnd}
\end{align}

Above integration is performed over the magnitude of momenta as well as their opening angle $\theta$. And if we fix the $\theta$, then it is convenient to trade $(s,|\mathbf{q}|)$ for the equal-momentum variables $(\mathbf{p},\theta)$, whose Jacobian at fixed $\theta$ is $|\partial(s,|\mathbf{q}|)/\partial(\mathbf{p},\theta)|=8\mathbf{p}^{2}\sin(\theta/2)$. The measure then becomes one-dimensional,
\begin{equation}
\bar P_{s\bar{s}}(\theta)=\frac{\displaystyle\int\!\mathrm{d}\mathbf{p}\;
\mathcal{J}\,W\,\mathcal{N}}{\displaystyle\int\!\mathrm{d}\mathbf{p}\;
\mathcal{J}\,W\,\mathcal{D}},
\,
\mathcal{J}=8\mathbf{p}^{2}\sin\frac{\theta}{2}\,\frac{\mathbf{q}^{2}}{2\omega}n_{B}(\omega)\mathcal{B}\beta(s).
\label{eq:Pbar}
\end{equation}

\subsection{Spin Correlation at $\mu_B=0$}
\label{subsec:zeromub}
\begin{figure*}[t]
\includegraphics[width=0.95\textwidth]{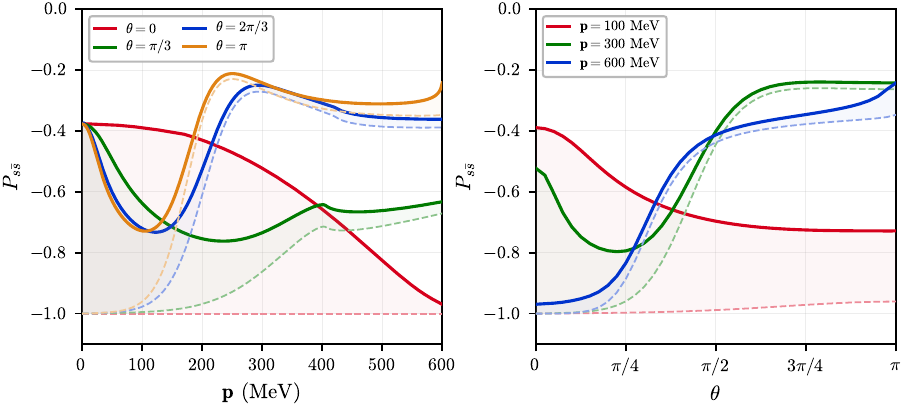}
\caption{Spin correlation $P_{s\bar{s}}(p,\theta)$ at  $T=T_{c}(0)=172.9$~MeV and $\mu_{B}=0$. (a) Take opening angles $\theta=0$, $\pi/3$, $2\pi/3$ and $\pi$, as a function of $\mathbf{p}$. (b) Fixed momenta $\mathbf{p}=100$, $300$ and $600$~MeV, as a function of the opening angle. In both panels the solid curves are $B_{0}=10$, dashed curves $B_{0}=0$.}
\label{fig:zeromub1}
\end{figure*}

We now work at vanishing chemical potential, $\mu_{B}=0$, and first consider the spin correlation exactly at the chiral crossover temperature of the NJL model, $T=T_{c}(0)=172.9$~MeV. The spin correlation is now a function of the momentum magnitude $\mathbf{p}$ and of the opening angle $\theta$, as shown in Fig.~\ref{fig:zeromub1}.

Fig.~\ref{fig:zeromub1}(a) shows $P_{s\bar{s}}$ as a function of momentum for
different opening angles $\theta$, where we plotted both case of $B_0=10$(solid lines) and  $B_0=0$(dashed lines), however, we concentrate on the more realistic $B_0=10$ case because that there must be much spin un-correlated quark pairs as background in a real HIC, and we also found that these two case differs significantly at low momentum while tend to equal for large momentum. At zero momentum limit, all $P_{s\bar{s}}$ has same value, regardless of the opening angle. This is understandable because now scalar meson drops out, and only pseudo-scalar meson contribute, which does not depend on $\theta$ as stated by Eq.~\eqref{eq:Pzerop}. When quark and anti-quark move in different direction, i.e., $\mathbf{p}>0$ and $\theta>0$, the spin correlation exhibits a non-monotonic behavior, with prominent local extrema emerging at specific angles, which is the manifestation of the competition between scalar and pseudo-scalar channel. And for the case that they move collinearly $\theta = 0$, $P_{s\bar{s}}$ decreases monotonically as momentum increases.

Fig.~\ref{fig:zeromub1}(b)  shows $P_{s\bar{s}}$ as a function of opening angle for given momentum: for low/large momentum($\mathbf{p}=100/600$MeV), $P_{s\bar{s}}$  increases/decreases with increasing $\theta$, while for medium momentum($\mathbf{p}=300$MeV), as shown by the green line,  $P_{s\bar{s}}$  arises first then decreases. Both figures show that $P_{s\bar{s}}$ could vary non-monotonically with momentum or opening angle induced by the competition of  scalar and pseudo-scalar channel.

\begin{figure*}[t]
\includegraphics[width=0.9\textwidth]{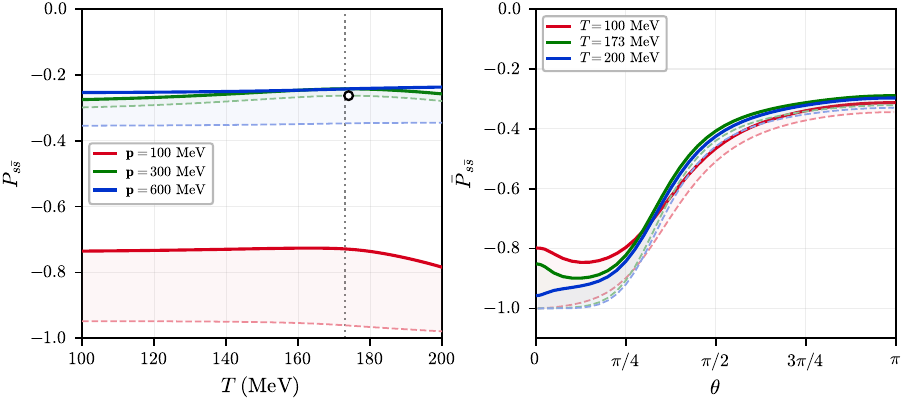}
\caption{(a) $P_{s\bar{s}}$ as a function of temperature at $\mu_{B}=0$ for the back-to-back pair $\theta=\pi$, the open circle marks the maximum of the $\mathbf{p}=300$~MeV curve and the vertical dotted line is $T_{c}(0)=172.9$~MeV. (b) The $\mathbf{p}$-integrated observable $\bar P_{s\bar{s}}$, as a function of the opening angle.}
\label{fig:zeromub2}
\end{figure*}

Next, we consider how spin correlation varies with temperature, as presented in Fig.~\ref{fig:zeromub2}. When quark and anti-quark move in opposite direction, Fig.~\ref{fig:zeromub2}(a) shows that at $\mathbf{p}=300$~MeV, $P_{s\bar{s}}$ rises from $-0.299$ at $T=100$~MeV to a maximum of $-0.263$ at $T=174$~MeV, which is essentially the crossover temperature, and then falls to $-0.279$ at $T=200$~MeV. The other two cases are monotone over this window. However, all three lines only change slightly with temperature, then it is impossible to identify the chiral phase transition directly from $P_{s\bar{s}}$. 

A more realistic quantity is the momentum-integrated spin correlation $\bar{P}_{s\bar{s}}$, with the measure of Eq.~\eqref{eq:Pbar}, as a function of opening angle, as shown in Fig.~\ref{fig:zeromub2} (b), where we choose three different temperature that are below, around and above $T_c(0)$, separately. All three lines show weaker spin correlation around $\theta=\pi$, where the quark and anti-quark move back-to-back, while the strongest correlation arises when they move almost in the same direction, $\theta\approx0$. It is interesting to note that, in $pp$ collisions, the momentum-averaged spin correlation is also found to be largest at small separation while spin correlation vanishes at large separation~\cite{STAR:2025njp}, which, however, exhibits positive sign of spin correlation.

\subsection{Spin Correlation Along the Freeze-out Line}
\label{subsec:freezeout}
\begin{figure}[t]
\includegraphics[width=0.95\columnwidth]{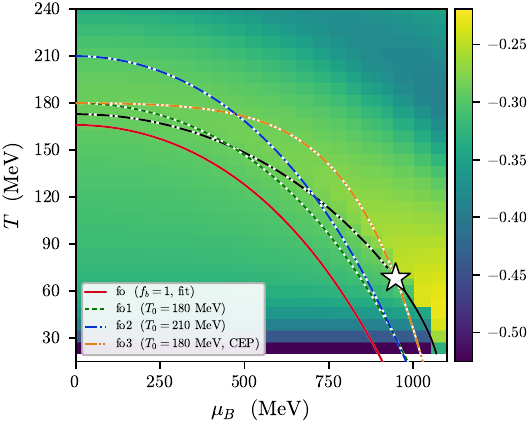}
\caption{$p$-integrated $P_{s\bar{s}}$ at $\theta=\pi$ and $B_{0}=10$ over the
$(T,\mu_{B})$ plane. The chiral crossover line (black dash-dotted) is drawn up to the critical end point, marked by the star at $(68.2,948)$~MeV; beyond it the transition is of first order and is continued as the black solid line down to $T=20$~MeV, obtained from the same strict coexistence criterion used to locate the end point. The four coloured curves are the chemical freeze-out trajectories of Eq.~\eqref{eq:freezeout}, distinguished by line style, labelled fo (solid red), fo1 (short-dashed green), fo2 (dash-dotted blue) and fo3 (dash-dot-dotted orange); they are the same lines, with the same coding, as in Fig.~\ref{fig:fopi}.}
\label{fig:phasediag}
\end{figure}

Now we consider the spin correlation  in the full $T-\mu_B$ plane, as shown in Fig.~\ref{fig:phasediag} by different colors. The opening angle is taken $\theta=\pi$, at which the scalar channel competes most strongly and also the natural configuration for a pair produced back-to-back from a nearly static source. And to connect to HIC experiment, it is more helpful to investigate how the spin correlation varies along freeze-out line, which fitted by experimental data has form of
\begin{align}
    \mu_{B}&=\frac{d}{1+e\sqrt{s_{NN}}},\quad T=a-b\,\mu_{B}^{2}-c\,\mu_{B}^{4},
    \label{eq:freezeout}
\end{align}
where $d=1.308$~GeV and $e=0.273$, and $a=0.166$~GeV, $b=0.139$~GeV$^{-1}$, $c=0.053$~GeV$^{-3}$~\cite{Cleymans:2005xv}. On the other hand, as we know that the crossover temperature obtained from the NJL model, $T_{c}(0)=172.9$~MeV in this paper, is higher than the results from LQCD, thus, besides the above one, labeled \texttt{fo}, we also choose another three freeze-out lines with labels \texttt{fo1}, \texttt{fo2} and \texttt{fo3}:
\begin{itemize}
\item \textbf{fo1}: $a=0.180$~GeV,   $b=0.119$~GeV$^{-1}$, then the curve starts at $T=180$~MeV at $\mu_{B}=0$, slightly above the chiral crossover $T_{c}(0)=172.9$~MeV;
\item \textbf{fo2}: $a=0.210$~GeV,  $b=0.153$~GeV$^{-1}$, starting at $T=210$~MeV at $\mu_{B}=0$;
\item \textbf{fo3}: $a=0.180$~GeV, $b=0.016$~GeV$^{-1}$, and add a sixth-order term $e_{6}\mu_{B}^{6}$ with $e_{6}=0.075$~GeV$^{-5}$,  and by these parameters, the curve passes exactly through the CEP.
\end{itemize}
All four freeze-out lines are shown in Fig.~\ref{fig:phasediag}, where the black dash-dotted line is the chiral crossover $T_{c}(\mu_{B})$ and the star marks the CEP at $(68.2,948)$~MeV.

\begin{figure}[t]
\includegraphics[width=0.95\columnwidth]{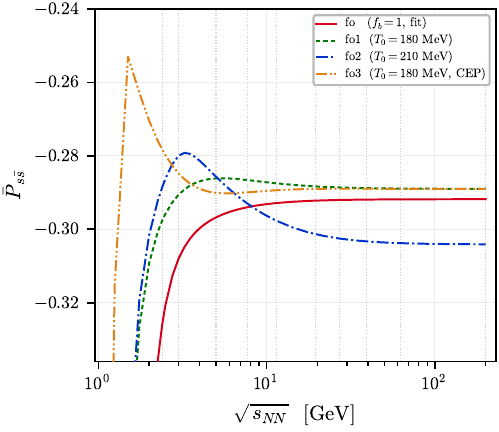}
\caption{$p$-integrated $P_{s\bar{s}}$ at $\theta=\pi$ along the four
freeze-out lines fo, fo1, fo2 and fo3 of Eq.~\eqref{eq:freezeout}, as a
function of collision energy; only the conservative $B_{0}=10$ curves are
shown.}
\label{fig:fopi}
\end{figure}

Fig.~\ref{fig:fopi} shows the result that the $\mathbf{p}$-integrated  spin correlation $\bar P_{s\bar{s}}$ as a function of $\sqrt{s_{NN}}$ along four different freeze-out lines. $\bar P_{s\bar{s}}$ along all four lines remain negative, which indicates that the pseudo-scalar channel dominates along every trajectory considered. Actually, it's clear from the Fig.~\ref{fig:phasediag}  that pseudo-scalar channel dominates all over the phase diagram. And as collision energy decreases, $\bar P_{s\bar{s}}$ along $fo$ decreases monotonically, while all $fo1 \sim fo3$ show non-monotonic behavior: $fo1$ has an interior minimum of $-0.286$ at $\sqrt{s_{NN}}\simeq5.4$~GeV, and $fo2$ reaching $-0.279$ at $\simeq3.3$~GeV. On the other hand, $\bar P_{s\bar{s}}$ along $fo3$ reaches minimum when it crosses CEP, which corresponds to $\sqrt{s_{NN}} \sim 2$GeV.

It should be noted that, because the observable samples invariant energies $\omega \ge 2M_s$, the low-lying resonance poles near the CEP are largely filtered out by the two-quark threshold. Consequently, the  spin correlation $P_{s\bar{s}}$ is not directly sensitive to localized critical fluctuations at the CEP itself. Instead, the non-monotonic behavior observed along trajectories $fo1\sim fo3$ arises because these trajectories cross the chiral crossover ridge $T \approx (1.03\text{--}1.09)T_c(\mu_B)$ in the $T\text{--}\mu_B$ plane. The emergence of the peak at $\sqrt{s_{NN}} \sim 2\text{--}5\text{ GeV}$ thus serves as an indicator of the freeze-out trajectory intersecting the chiral phase boundary, rather than a direct critical fluctuation signal of the CEP.
\section{Conclusions}
\label{sec:conclusions}

We have studied how the spin correlation of $\Lambda\bar{\Lambda}$ pairs is built up from the two mesonic channels that mediate $s\bar{s}$ production in the medium. Using Yukawa-like interaction, we derived the spin correlation $P_{s\bar{s}}$ from the contribution of scalar and pseudo-scalar channel, and it is found that the scalar and pseudo-scalar channels  project the pair onto different total-spin states, ${}^{3}P_{0}$ and ${}^{1}S_{0}$, which contribute opposite spin correlations, $+1/3$ and $-1$: the scalar and pseudo-scalar channels thus compete with each other.

To make quantitative prediction with dynamical quarks, we have used 3-flavor NJL model with RPA approximation to extract the meson spectral in the $T-\mu_B$ plane, and found that the pseudo-scalar channel dominates and  $\bar P_{s\bar{s}}<0$ throughout all the accessible region. The suppression of the scalar channel results from the combined effect of the $1{:}3$ spin weight and the $P$-wave barrier $\mathcal{P}_{-}/\mathcal{P}_{+}=1-4M_{s}^{2}/s<1$. 

At vanishing chemical potential, we found $P_{s\bar{s}}$ rarely change with temperature, not to mention non-monotonic behavior. Although strange quark sector can feel chiral phase transition through the coupling with light quarks, the  mass of strange meson does not change significantly around chiral crossover temperature, as well as that the coupling tends to vanish at chiral symmetry restoration phase. However, $\bar{P}_{s\bar{s}}$ changes significantly with momentum as well as the opening angle, and the momentum-integrated $\bar P_{s\bar{s}}$ as a function of opening angle shows strongest correlation around $\theta=0$, while the weakest correlation at  $\theta=\pi$.

More interesting results is about the momentum-integrated $\bar P_{s\bar{s}}$ at $\theta=\pi$ along freeze-out lines. And it is found that $\bar P_{s\bar{s}}$ along the freeze-out line fitted by experiment data shows less interesting behavior, however, a non-monotonic behavior emerges when an appropriate freeze-out line is chosen, which starts above the crossover temperature at $\mu_B=0$ and then cross the phase boundary line before reaching CEP. The resulting minimum near $\sqrt{s_{NN}}\simeq2$--$5$~GeV is a signature of the trajectory crossing the chiral crossover, and get enhanced as the intersection moving to CEP, within reach of the RHIC BES and the low-energy programmes at FAIR and HADES~\cite{STAR:2021fge,CBM:2016kpk,HADES:2020wpc} at  $\sqrt{s_{NN}}$ around $2\sim 5$~GeV.

\begin{acknowledgments}
This work is supported  by the National Natural Science Foundation of China (NSFC) Grant No.12505148.
\end{acknowledgments}

\bibliography{reference}

\begin{thebibliography}{34}%
\makeatletter
\providecommand \@ifxundefined [1]{%
 \@ifx{#1\undefined}
}%
\providecommand \@ifnum [1]{%
 \ifnum #1\expandafter \@firstoftwo
 \else \expandafter \@secondoftwo
 \fi
}%
\providecommand \@ifx [1]{%
 \ifx #1\expandafter \@firstoftwo
 \else \expandafter \@secondoftwo
 \fi
}%
\providecommand \natexlab [1]{#1}%
\providecommand \enquote  [1]{``#1''}%
\providecommand \bibnamefont  [1]{#1}%
\providecommand \bibfnamefont [1]{#1}%
\providecommand \citenamefont [1]{#1}%
\providecommand \href@noop [0]{\@secondoftwo}%
\providecommand \href [0]{\begingroup \@sanitize@url \@href}%
\providecommand \@href[1]{\@@startlink{#1}\@@href}%
\providecommand \@@href[1]{\endgroup#1\@@endlink}%
\providecommand \@sanitize@url [0]{\catcode `\\12\catcode `\$12\catcode
  `\&12\catcode `\#12\catcode `\^12\catcode `\_12\catcode `\%12\relax}%
\providecommand \@@startlink[1]{}%
\providecommand \@@endlink[0]{}%
\providecommand \url  [0]{\begingroup\@sanitize@url \@url }%
\providecommand \@url [1]{\endgroup\@href {#1}{\urlprefix }}%
\providecommand \urlprefix  [0]{URL }%
\providecommand \Eprint [0]{\href }%
\providecommand \doibase [0]{https://doi.org/}%
\providecommand \selectlanguage [0]{\@gobble}%
\providecommand \bibinfo  [0]{\@secondoftwo}%
\providecommand \bibfield  [0]{\@secondoftwo}%
\providecommand \translation [1]{[#1]}%
\providecommand \BibitemOpen [0]{}%
\providecommand \bibitemStop [0]{}%
\providecommand \bibitemNoStop [0]{.\EOS\space}%
\providecommand \EOS [0]{\spacefactor3000\relax}%
\providecommand \BibitemShut  [1]{\csname bibitem#1\endcsname}%
\let\auto@bib@innerbib\@empty
\bibitem [{\citenamefont {Adamczyk}\ \emph {et~al.}(2017)\citenamefont
  {Adamczyk} \emph {et~al.}}]{STAR:2017ckg}%
  \BibitemOpen
  \bibfield  {author} {\bibinfo {author} {\bibfnamefont {L.}~\bibnamefont
  {Adamczyk}} \emph {et~al.} (\bibinfo {collaboration} {STAR}),\ }\bibfield
  {title} {\bibinfo {title} {{Global $\Lambda$ hyperon polarization in nuclear
  collisions: evidence for the most vortical fluid}},\ }\href
  {https://doi.org/10.1038/nature23004} {\bibfield  {journal} {\bibinfo
  {journal} {Nature}\ }\textbf {\bibinfo {volume} {548}},\ \bibinfo {pages}
  {62} (\bibinfo {year} {2017})},\ \Eprint {https://arxiv.org/abs/1701.06657}
  {arXiv:1701.06657 [nucl-ex]} \BibitemShut {NoStop}%
\bibitem [{\citenamefont {Abdallah}\ \emph {et~al.}(2021)\citenamefont
  {Abdallah} \emph {et~al.}}]{STAR:2021beb}%
  \BibitemOpen
  \bibfield  {author} {\bibinfo {author} {\bibfnamefont {M.~S.}\ \bibnamefont
  {Abdallah}} \emph {et~al.} (\bibinfo {collaboration} {STAR}),\ }\bibfield
  {title} {\bibinfo {title} {{Global $\Lambda$-hyperon polarization in Au+Au
  collisions at $\sqrt {s_{NN}}$=3~GeV}},\ }\href
  {https://doi.org/10.1103/PhysRevC.104.L061901} {\bibfield  {journal}
  {\bibinfo  {journal} {Phys. Rev. C}\ }\textbf {\bibinfo {volume} {104}},\
  \bibinfo {pages} {L061901} (\bibinfo {year} {2021})},\ \Eprint
  {https://arxiv.org/abs/2108.00044} {arXiv:2108.00044 [nucl-ex]} \BibitemShut
  {NoStop}%
\bibitem [{\citenamefont {Adam}\ \emph {et~al.}(2021)\citenamefont {Adam} \emph
  {et~al.}}]{STAR:2020xbm}%
  \BibitemOpen
  \bibfield  {author} {\bibinfo {author} {\bibfnamefont {J.}~\bibnamefont
  {Adam}} \emph {et~al.} (\bibinfo {collaboration} {STAR}),\ }\bibfield
  {title} {\bibinfo {title} {{Global Polarization of $\Xi$ and $\Omega$
  Hyperons in Au+Au Collisions at $\sqrt {s_{NN}}$ = 200 GeV}},\ }\href
  {https://doi.org/10.1103/PhysRevLett.126.162301} {\bibfield  {journal}
  {\bibinfo  {journal} {Phys. Rev. Lett.}\ }\textbf {\bibinfo {volume} {126}},\
  \bibinfo {pages} {162301} (\bibinfo {year} {2021})},\ \bibinfo {note}
  {[Erratum: Phys.Rev.Lett. 131, 089901 (2023)]},\ \Eprint
  {https://arxiv.org/abs/2012.13601} {arXiv:2012.13601 [nucl-ex]} \BibitemShut
  {NoStop}%
\bibitem [{\citenamefont {Becattini}\ \emph {et~al.}(2013)\citenamefont
  {Becattini}, \citenamefont {Chandra}, \citenamefont {Del~Zanna},\ and\
  \citenamefont {Grossi}}]{Becattini:2013fla}%
  \BibitemOpen
  \bibfield  {author} {\bibinfo {author} {\bibfnamefont {F.}~\bibnamefont
  {Becattini}}, \bibinfo {author} {\bibfnamefont {V.}~\bibnamefont {Chandra}},
  \bibinfo {author} {\bibfnamefont {L.}~\bibnamefont {Del~Zanna}},\ and\
  \bibinfo {author} {\bibfnamefont {E.}~\bibnamefont {Grossi}},\ }\bibfield
  {title} {\bibinfo {title} {{Relativistic distribution function for particles
  with spin at local thermodynamical equilibrium}},\ }\href
  {https://doi.org/10.1016/j.aop.2013.07.004} {\bibfield  {journal} {\bibinfo
  {journal} {Annals Phys.}\ }\textbf {\bibinfo {volume} {338}},\ \bibinfo
  {pages} {32} (\bibinfo {year} {2013})},\ \Eprint
  {https://arxiv.org/abs/1303.3431} {arXiv:1303.3431 [nucl-th]} \BibitemShut
  {NoStop}%
\bibitem [{\citenamefont {Karpenko}\ and\ \citenamefont
  {Becattini}(2017)}]{Karpenko:2016jyx}%
  \BibitemOpen
  \bibfield  {author} {\bibinfo {author} {\bibfnamefont {I.}~\bibnamefont
  {Karpenko}}\ and\ \bibinfo {author} {\bibfnamefont {F.}~\bibnamefont
  {Becattini}},\ }\bibfield  {title} {\bibinfo {title} {{Study of $\Lambda$
  polarization in relativistic nuclear collisions at
  $\sqrt{s_\mathrm{NN}}=7.7$--200 GeV}},\ }\href
  {https://doi.org/10.1140/epjc/s10052-017-4765-1} {\bibfield  {journal}
  {\bibinfo  {journal} {Eur. Phys. J. C}\ }\textbf {\bibinfo {volume} {77}},\
  \bibinfo {pages} {213} (\bibinfo {year} {2017})},\ \Eprint
  {https://arxiv.org/abs/1610.04717} {arXiv:1610.04717 [nucl-th]} \BibitemShut
  {NoStop}%
\bibitem [{\citenamefont {Xu}\ \emph {et~al.}(2022)\citenamefont {Xu},
  \citenamefont {Lin}, \citenamefont {Huang},\ and\ \citenamefont
  {Huang}}]{Xu:2022hql}%
  \BibitemOpen
  \bibfield  {author} {\bibinfo {author} {\bibfnamefont {K.}~\bibnamefont
  {Xu}}, \bibinfo {author} {\bibfnamefont {F.}~\bibnamefont {Lin}}, \bibinfo
  {author} {\bibfnamefont {A.}~\bibnamefont {Huang}},\ and\ \bibinfo {author}
  {\bibfnamefont {M.}~\bibnamefont {Huang}},\ }\bibfield  {title} {\bibinfo
  {title} {{\ensuremath{\Lambda}/\ensuremath{\Lambda}\textasciimacron{}
  polarization and splitting induced by rotation and magnetic field}},\ }\href
  {https://doi.org/10.1103/PhysRevD.106.L071502} {\bibfield  {journal}
  {\bibinfo  {journal} {Phys. Rev. D}\ }\textbf {\bibinfo {volume} {106}},\
  \bibinfo {pages} {L071502} (\bibinfo {year} {2022})},\ \Eprint
  {https://arxiv.org/abs/2205.02420} {arXiv:2205.02420 [hep-ph]} \BibitemShut
  {NoStop}%
\bibitem [{\citenamefont {Huang}(2020)}]{Huang:2020vsp}%
  \BibitemOpen
  \bibfield  {author} {\bibinfo {author} {\bibfnamefont {X.-G.}\ \bibnamefont
  {Huang}},\ }\bibfield  {title} {\bibinfo {title} {{Vorticity and Spin
  Polarization --- A Theoretical Perspective}},\ }\href
  {https://doi.org/10.1016/j.nuclphysa.2020.121752} {\bibfield  {journal}
  {\bibinfo  {journal} {Nucl. Phys. A}\ }\textbf {\bibinfo {volume} {1001}},\
  \bibinfo {pages} {121752} (\bibinfo {year} {2020})},\ \Eprint
  {https://arxiv.org/abs/2002.07549} {arXiv:2002.07549 [nucl-th]} \BibitemShut
  {NoStop}%
\bibitem [{\citenamefont {Acharya}\ \emph {et~al.}(2020)\citenamefont {Acharya}
  \emph {et~al.}}]{ALICE:2019aid}%
  \BibitemOpen
  \bibfield  {author} {\bibinfo {author} {\bibfnamefont {S.}~\bibnamefont
  {Acharya}} \emph {et~al.} (\bibinfo {collaboration} {ALICE}),\ }\bibfield
  {title} {\bibinfo {title} {{Evidence of Spin-Orbital Angular Momentum
  Interactions in Relativistic Heavy-Ion Collisions}},\ }\href
  {https://doi.org/10.1103/PhysRevLett.125.012301} {\bibfield  {journal}
  {\bibinfo  {journal} {Phys. Rev. Lett.}\ }\textbf {\bibinfo {volume} {125}},\
  \bibinfo {pages} {012301} (\bibinfo {year} {2020})},\ \Eprint
  {https://arxiv.org/abs/1910.14408} {arXiv:1910.14408 [nucl-ex]} \BibitemShut
  {NoStop}%
\bibitem [{\citenamefont {Abdallah}\ \emph {et~al.}(2023)\citenamefont
  {Abdallah} \emph {et~al.}}]{STAR:2022fan}%
  \BibitemOpen
  \bibfield  {author} {\bibinfo {author} {\bibfnamefont {M.~S.}\ \bibnamefont
  {Abdallah}} \emph {et~al.} (\bibinfo {collaboration} {STAR}),\ }\bibfield
  {title} {\bibinfo {title} {{Pattern of global spin alignment of $\phi$ and
  $K^{*0}$ mesons in heavy-ion collisions}},\ }\href
  {https://doi.org/10.1038/s41586-022-05557-5} {\bibfield  {journal} {\bibinfo
  {journal} {Nature}\ }\textbf {\bibinfo {volume} {614}},\ \bibinfo {pages}
  {244} (\bibinfo {year} {2023})},\ \Eprint {https://arxiv.org/abs/2204.02302}
  {arXiv:2204.02302 [hep-ph]} \BibitemShut {NoStop}%
\bibitem [{\citenamefont {Wei}\ and\ \citenamefont
  {Huang}(2023)}]{Wei:2023pdf}%
  \BibitemOpen
  \bibfield  {author} {\bibinfo {author} {\bibfnamefont {M.}~\bibnamefont
  {Wei}}\ and\ \bibinfo {author} {\bibfnamefont {M.}~\bibnamefont {Huang}},\
  }\bibfield  {title} {\bibinfo {title} {{Spin alignment of vector mesons from
  quark dynamics in a rotating medium}},\ }\href@noop {} {\  (\bibinfo {year}
  {2023})},\ \Eprint {https://arxiv.org/abs/2303.01897} {arXiv:2303.01897
  [hep-ph]} \BibitemShut {NoStop}%
\bibitem [{\citenamefont {Sheng}\ \emph {et~al.}(2022)\citenamefont {Sheng},
  \citenamefont {Yang}, \citenamefont {Zou},\ and\ \citenamefont
  {Hou}}]{Sheng:2022ssp}%
  \BibitemOpen
  \bibfield  {author} {\bibinfo {author} {\bibfnamefont {X.-L.}\ \bibnamefont
  {Sheng}}, \bibinfo {author} {\bibfnamefont {S.-Y.}\ \bibnamefont {Yang}},
  \bibinfo {author} {\bibfnamefont {Y.-L.}\ \bibnamefont {Zou}},\ and\ \bibinfo
  {author} {\bibfnamefont {D.}~\bibnamefont {Hou}},\ }\bibfield  {title}
  {\bibinfo {title} {{Mass splitting and spin alignment for $\phi$ mesons in a
  magnetic field in NJL model}},\ }\href@noop {} {\  (\bibinfo {year}
  {2022})},\ \Eprint {https://arxiv.org/abs/2209.01872} {arXiv:2209.01872
  [nucl-th]} \BibitemShut {NoStop}%
\bibitem [{\citenamefont {Acharya}\ \emph {et~al.}(2023)\citenamefont {Acharya}
  \emph {et~al.}}]{ALICE:2022dyy}%
  \BibitemOpen
  \bibfield  {author} {\bibinfo {author} {\bibfnamefont {S.}~\bibnamefont
  {Acharya}} \emph {et~al.} (\bibinfo {collaboration} {ALICE}),\ }\bibfield
  {title} {\bibinfo {title} {{Measurement of the J/{\ensuremath{\psi}}
  Polarization with Respect to the Event Plane in Pb-Pb Collisions at the
  LHC}},\ }\href {https://doi.org/10.1103/PhysRevLett.131.042303} {\bibfield
  {journal} {\bibinfo  {journal} {Phys. Rev. Lett.}\ }\textbf {\bibinfo
  {volume} {131}},\ \bibinfo {pages} {042303} (\bibinfo {year} {2023})},\
  \Eprint {https://arxiv.org/abs/2204.10171} {arXiv:2204.10171 [nucl-ex]}
  \BibitemShut {NoStop}%
\bibitem [{\citenamefont {Yang}\ \emph {et~al.}(2018)\citenamefont {Yang},
  \citenamefont {Fang}, \citenamefont {Wang},\ and\ \citenamefont
  {Wang}}]{Yang:2017sdk}%
  \BibitemOpen
  \bibfield  {author} {\bibinfo {author} {\bibfnamefont {Y.-G.}\ \bibnamefont
  {Yang}}, \bibinfo {author} {\bibfnamefont {R.-H.}\ \bibnamefont {Fang}},
  \bibinfo {author} {\bibfnamefont {Q.}~\bibnamefont {Wang}},\ and\ \bibinfo
  {author} {\bibfnamefont {X.-N.}\ \bibnamefont {Wang}},\ }\bibfield  {title}
  {\bibinfo {title} {{Quark coalescence model for polarized vector mesons and
  baryons}},\ }\href {https://doi.org/10.1103/PhysRevC.97.034917} {\bibfield
  {journal} {\bibinfo  {journal} {Phys. Rev. C}\ }\textbf {\bibinfo {volume}
  {97}},\ \bibinfo {pages} {034917} (\bibinfo {year} {2018})},\ \Eprint
  {https://arxiv.org/abs/1711.06008} {arXiv:1711.06008 [nucl-th]} \BibitemShut
  {NoStop}%
\bibitem [{\citenamefont {Xia}\ \emph {et~al.}(2021)\citenamefont {Xia},
  \citenamefont {Li}, \citenamefont {Huang},\ and\ \citenamefont
  {Zhong~Huang}}]{Xia:2020tyd}%
  \BibitemOpen
  \bibfield  {author} {\bibinfo {author} {\bibfnamefont {X.-L.}\ \bibnamefont
  {Xia}}, \bibinfo {author} {\bibfnamefont {H.}~\bibnamefont {Li}}, \bibinfo
  {author} {\bibfnamefont {X.-G.}\ \bibnamefont {Huang}},\ and\ \bibinfo
  {author} {\bibfnamefont {H.}~\bibnamefont {Zhong~Huang}},\ }\bibfield
  {title} {\bibinfo {title} {{Local spin alignment of vector mesons in
  relativistic heavy-ion collisions}},\ }\href
  {https://doi.org/10.1016/j.physletb.2021.136325} {\bibfield  {journal}
  {\bibinfo  {journal} {Phys. Lett. B}\ }\textbf {\bibinfo {volume} {817}},\
  \bibinfo {pages} {136325} (\bibinfo {year} {2021})},\ \Eprint
  {https://arxiv.org/abs/2010.01474} {arXiv:2010.01474 [nucl-th]} \BibitemShut
  {NoStop}%
\bibitem [{\citenamefont {Sheng}\ \emph {et~al.}(2020)\citenamefont {Sheng},
  \citenamefont {Oliva},\ and\ \citenamefont {Wang}}]{Sheng:2019kmk}%
  \BibitemOpen
  \bibfield  {author} {\bibinfo {author} {\bibfnamefont {X.-L.}\ \bibnamefont
  {Sheng}}, \bibinfo {author} {\bibfnamefont {L.}~\bibnamefont {Oliva}},\ and\
  \bibinfo {author} {\bibfnamefont {Q.}~\bibnamefont {Wang}},\ }\bibfield
  {title} {\bibinfo {title} {{What can we learn from the global spin alignment
  of $\phi$ mesons in heavy-ion collisions?}},\ }\href
  {https://doi.org/10.1103/PhysRevD.101.096005} {\bibfield  {journal} {\bibinfo
   {journal} {Phys. Rev. D}\ }\textbf {\bibinfo {volume} {101}},\ \bibinfo
  {pages} {096005} (\bibinfo {year} {2020})},\ \bibinfo {note} {[Erratum: Phys.
  Rev. D 105, 099903 (2022)]},\ \Eprint {https://arxiv.org/abs/1910.13684}
  {arXiv:1910.13684 [nucl-th]} \BibitemShut {NoStop}%
\bibitem [{\citenamefont {Sheng}\ \emph
  {et~al.}(2023{\natexlab{a}})\citenamefont {Sheng}, \citenamefont {Oliva},
  \citenamefont {Liang}, \citenamefont {Wang},\ and\ \citenamefont
  {Wang}}]{Sheng:2022sva}%
  \BibitemOpen
  \bibfield  {author} {\bibinfo {author} {\bibfnamefont {X.-L.}\ \bibnamefont
  {Sheng}}, \bibinfo {author} {\bibfnamefont {L.}~\bibnamefont {Oliva}},
  \bibinfo {author} {\bibfnamefont {Z.-T.}\ \bibnamefont {Liang}}, \bibinfo
  {author} {\bibfnamefont {Q.}~\bibnamefont {Wang}},\ and\ \bibinfo {author}
  {\bibfnamefont {X.-N.}\ \bibnamefont {Wang}},\ }\bibfield  {title} {\bibinfo
  {title} {{Spin alignment of vector mesons in heavy-ion collisions}},\ }\href
  {https://doi.org/10.1103/PhysRevLett.131.042304} {\bibfield  {journal}
  {\bibinfo  {journal} {Phys. Rev. Lett.}\ }\textbf {\bibinfo {volume} {131}},\
  \bibinfo {pages} {042304} (\bibinfo {year} {2023}{\natexlab{a}})},\ \Eprint
  {https://arxiv.org/abs/2205.15689} {arXiv:2205.15689 [hep-ph]} \BibitemShut
  {NoStop}%
\bibitem [{\citenamefont {Sheng}\ \emph
  {et~al.}(2023{\natexlab{b}})\citenamefont {Sheng}, \citenamefont {Pu},\ and\
  \citenamefont {Wang}}]{Sheng:2023urn}%
  \BibitemOpen
  \bibfield  {author} {\bibinfo {author} {\bibfnamefont {X.-L.}\ \bibnamefont
  {Sheng}}, \bibinfo {author} {\bibfnamefont {S.}~\bibnamefont {Pu}},\ and\
  \bibinfo {author} {\bibfnamefont {Q.}~\bibnamefont {Wang}},\ }\bibfield
  {title} {\bibinfo {title} {{Momentum dependence of $\phi$ meson's spin
  alignment}},\ }\href@noop {} {\  (\bibinfo {year} {2023}{\natexlab{b}})},\
  \Eprint {https://arxiv.org/abs/2308.14038} {arXiv:2308.14038 [nucl-th]}
  \BibitemShut {NoStop}%
\bibitem [{\citenamefont {Sheng}\ \emph {et~al.}(2026)\citenamefont {Sheng},
  \citenamefont {Wu}, \citenamefont {Rischke},\ and\ \citenamefont
  {Wang}}]{Sheng:2025puj}%
  \BibitemOpen
  \bibfield  {author} {\bibinfo {author} {\bibfnamefont {X.-L.}\ \bibnamefont
  {Sheng}}, \bibinfo {author} {\bibfnamefont {X.-Y.}\ \bibnamefont {Wu}},
  \bibinfo {author} {\bibfnamefont {D.~H.}\ \bibnamefont {Rischke}},\ and\
  \bibinfo {author} {\bibfnamefont {X.-N.}\ \bibnamefont {Wang}},\ }\bibfield
  {title} {\bibinfo {title} {{Hyperon Spin Correlation in High-Energy Heavy-Ion
  Collisions}},\ }\href {https://doi.org/10.1103/z5kk-98my} {\bibfield
  {journal} {\bibinfo  {journal} {Phys. Rev. Lett.}\ }\textbf {\bibinfo
  {volume} {136}},\ \bibinfo {pages} {082301} (\bibinfo {year} {2026})},\
  \Eprint {https://arxiv.org/abs/2508.03496} {arXiv:2508.03496 [hep-ph]}
  \BibitemShut {NoStop}%
\bibitem [{\citenamefont {M{\"u}ller}\ and\ \citenamefont
  {Yang}(2022)}]{Muller:2022tcv}%
  \BibitemOpen
  \bibfield  {author} {\bibinfo {author} {\bibfnamefont {B.}~\bibnamefont
  {M{\"u}ller}}\ and\ \bibinfo {author} {\bibfnamefont {D.-L.}\ \bibnamefont
  {Yang}},\ }\bibfield  {title} {\bibinfo {title} {{Spin alignment from
  turbulent color fields}},\ }\href@noop {} {\  (\bibinfo {year} {2022})},\
  \Eprint {https://arxiv.org/abs/2207.14504} {arXiv:2207.14504 [nucl-th]}
  \BibitemShut {NoStop}%
\bibitem [{\citenamefont {Chen}\ \emph {et~al.}(2023)\citenamefont {Chen},
  \citenamefont {Liang}, \citenamefont {Ma},\ and\ \citenamefont
  {Wang}}]{Chen:2023sfa}%
  \BibitemOpen
  \bibfield  {author} {\bibinfo {author} {\bibfnamefont {J.}~\bibnamefont
  {Chen}}, \bibinfo {author} {\bibfnamefont {Z.-T.}\ \bibnamefont {Liang}},
  \bibinfo {author} {\bibfnamefont {Y.-G.}\ \bibnamefont {Ma}},\ and\ \bibinfo
  {author} {\bibfnamefont {Q.}~\bibnamefont {Wang}},\ }\bibfield  {title}
  {\bibinfo {title} {{Global spin alignment of vector mesons and strong force
  fields in heavy-ion collisions}},\ }\href
  {https://doi.org/10.1016/j.scib.2023.04.001} {\bibfield  {journal} {\bibinfo
  {journal} {Sci. Bull.}\ }\textbf {\bibinfo {volume} {68}},\ \bibinfo {pages}
  {874} (\bibinfo {year} {2023})},\ \Eprint {https://arxiv.org/abs/2305.09114}
  {arXiv:2305.09114 [nucl-th]} \BibitemShut {NoStop}%
\bibitem [{\citenamefont {Xu}\ and\ \citenamefont {Huang}(2024)}]{Xu:2024kdh}%
  \BibitemOpen
  \bibfield  {author} {\bibinfo {author} {\bibfnamefont {K.}~\bibnamefont
  {Xu}}\ and\ \bibinfo {author} {\bibfnamefont {M.}~\bibnamefont {Huang}},\
  }\bibfield  {title} {\bibinfo {title} {{Spin alignment of vector mesons
  induced by local spin density fluctuations}},\ }\href
  {https://doi.org/10.1103/PhysRevD.110.094034} {\bibfield  {journal} {\bibinfo
   {journal} {Phys. Rev. D}\ }\textbf {\bibinfo {volume} {110}},\ \bibinfo
  {pages} {094034} (\bibinfo {year} {2024})},\ \Eprint
  {https://arxiv.org/abs/2408.06581} {arXiv:2408.06581 [hep-ph]} \BibitemShut
  {NoStop}%
\bibitem [{\citenamefont {Aboona}\ \emph {et~al.}(2026)\citenamefont {Aboona}
  \emph {et~al.}}]{STAR:2025njp}%
  \BibitemOpen
  \bibfield  {author} {\bibinfo {author} {\bibfnamefont {B.~E.}\ \bibnamefont
  {Aboona}} \emph {et~al.} (\bibinfo {collaboration} {STAR}),\ }\bibfield
  {title} {\bibinfo {title} {{Measuring spin correlation between quarks during
  QCD confinement}},\ }\href {https://doi.org/10.1038/s41586-025-09920-0}
  {\bibfield  {journal} {\bibinfo  {journal} {Nature}\ }\textbf {\bibinfo
  {volume} {650}},\ \bibinfo {pages} {65} (\bibinfo {year} {2026})},\ \Eprint
  {https://arxiv.org/abs/2506.05499} {arXiv:2506.05499 [hep-ex]} \BibitemShut
  {NoStop}%
\bibitem [{\citenamefont {Chen}\ \emph {et~al.}(2025)\citenamefont {Chen},
  \citenamefont {Fu}, \citenamefont {Huang},\ and\ \citenamefont
  {Ma}}]{Chen:2024hki}%
  \BibitemOpen
  \bibfield  {author} {\bibinfo {author} {\bibfnamefont {H.-L.}\ \bibnamefont
  {Chen}}, \bibinfo {author} {\bibfnamefont {W.-j.}\ \bibnamefont {Fu}},
  \bibinfo {author} {\bibfnamefont {X.-G.}\ \bibnamefont {Huang}},\ and\
  \bibinfo {author} {\bibfnamefont {G.-L.}\ \bibnamefont {Ma}},\ }\bibfield
  {title} {\bibinfo {title} {{Fluctuations and Correlations of Quark Spin in
  Hot and Dense QCD Matter}},\ }\href {https://doi.org/10.1103/g1bh-85h4}
  {\bibfield  {journal} {\bibinfo  {journal} {Phys. Rev. Lett.}\ }\textbf
  {\bibinfo {volume} {135}},\ \bibinfo {pages} {032302} (\bibinfo {year}
  {2025})},\ \Eprint {https://arxiv.org/abs/2410.20704} {arXiv:2410.20704
  [hep-ph]} \BibitemShut {NoStop}%
\bibitem [{\citenamefont {Lv}\ \emph {et~al.}(2024)\citenamefont {Lv},
  \citenamefont {Yu}, \citenamefont {Liang}, \citenamefont {Wang},\ and\
  \citenamefont {Wang}}]{Lv:2024uev}%
  \BibitemOpen
  \bibfield  {author} {\bibinfo {author} {\bibfnamefont {J.-p.}\ \bibnamefont
  {Lv}}, \bibinfo {author} {\bibfnamefont {Z.-h.}\ \bibnamefont {Yu}}, \bibinfo
  {author} {\bibfnamefont {Z.-t.}\ \bibnamefont {Liang}}, \bibinfo {author}
  {\bibfnamefont {Q.}~\bibnamefont {Wang}},\ and\ \bibinfo {author}
  {\bibfnamefont {X.-N.}\ \bibnamefont {Wang}},\ }\bibfield  {title} {\bibinfo
  {title} {{Global quark spin correlations in relativistic heavy ion
  collisions}},\ }\href {https://doi.org/10.1103/PhysRevD.109.114003}
  {\bibfield  {journal} {\bibinfo  {journal} {Phys. Rev. D}\ }\textbf {\bibinfo
  {volume} {109}},\ \bibinfo {pages} {114003} (\bibinfo {year} {2024})},\
  \Eprint {https://arxiv.org/abs/2402.13721} {arXiv:2402.13721 [hep-ph]}
  \BibitemShut {NoStop}%
\bibitem [{\citenamefont {Oliva}\ \emph {et~al.}(2026)\citenamefont {Oliva},
  \citenamefont {Wang},\ and\ \citenamefont {Wang}}]{Oliwa:2026qsc}%
  \BibitemOpen
  \bibfield  {author} {\bibinfo {author} {\bibfnamefont {L.}~\bibnamefont
  {Oliva}}, \bibinfo {author} {\bibfnamefont {Q.}~\bibnamefont {Wang}},\ and\
  \bibinfo {author} {\bibfnamefont {X.-N.}\ \bibnamefont {Wang}},\ }\bibfield
  {title} {\bibinfo {title} {{Quark spin correlation inside hyperons}},\
  }\href@noop {} {\  (\bibinfo {year} {2026})},\ \Eprint
  {https://arxiv.org/abs/2603.10427} {arXiv:2603.10427 [hep-ph]} \BibitemShut
  {NoStop}%
\bibitem [{\citenamefont {Nambu}\ and\ \citenamefont
  {Jona-Lasinio}(1961)}]{Nambu:1961tp}%
  \BibitemOpen
  \bibfield  {author} {\bibinfo {author} {\bibfnamefont {Y.}~\bibnamefont
  {Nambu}}\ and\ \bibinfo {author} {\bibfnamefont {G.}~\bibnamefont
  {Jona-Lasinio}},\ }\bibfield  {title} {\bibinfo {title} {{Dynamical Model of
  Elementary Particles Based on an Analogy with Superconductivity. 1.}},\
  }\href {https://doi.org/10.1103/PhysRev.122.345} {\bibfield  {journal}
  {\bibinfo  {journal} {Phys. Rev.}\ }\textbf {\bibinfo {volume} {122}},\
  \bibinfo {pages} {345} (\bibinfo {year} {1961})}\BibitemShut {NoStop}%
\bibitem [{\citenamefont {Klevansky}(1992)}]{Klevansky:1992qe}%
  \BibitemOpen
  \bibfield  {author} {\bibinfo {author} {\bibfnamefont {S.~P.}\ \bibnamefont
  {Klevansky}},\ }\bibfield  {title} {\bibinfo {title} {{The Nambu-Jona-Lasinio
  model of quantum chromodynamics}},\ }\href
  {https://doi.org/10.1103/RevModPhys.64.649} {\bibfield  {journal} {\bibinfo
  {journal} {Rev. Mod. Phys.}\ }\textbf {\bibinfo {volume} {64}},\ \bibinfo
  {pages} {649} (\bibinfo {year} {1992})}\BibitemShut {NoStop}%
\bibitem [{\citenamefont {Hatsuda}\ and\ \citenamefont
  {Kunihiro}(1994)}]{Hatsuda:1994pi}%
  \BibitemOpen
  \bibfield  {author} {\bibinfo {author} {\bibfnamefont {T.}~\bibnamefont
  {Hatsuda}}\ and\ \bibinfo {author} {\bibfnamefont {T.}~\bibnamefont
  {Kunihiro}},\ }\bibfield  {title} {\bibinfo {title} {{QCD phenomenology based
  on a chiral effective Lagrangian}},\ }\href
  {https://doi.org/10.1016/0370-1573(94)90022-1} {\bibfield  {journal}
  {\bibinfo  {journal} {Phys. Rept.}\ }\textbf {\bibinfo {volume} {247}},\
  \bibinfo {pages} {221} (\bibinfo {year} {1994})},\ \Eprint
  {https://arxiv.org/abs/hep-ph/9401310} {arXiv:hep-ph/9401310 [hep-ph]}
  \BibitemShut {NoStop}%
\bibitem [{\citenamefont {Hansen}\ \emph {et~al.}(2007)\citenamefont {Hansen},
  \citenamefont {Alberico}, \citenamefont {Beraudo}, \citenamefont {Molinari},
  \citenamefont {Nardi},\ and\ \citenamefont {Ratti}}]{Hansen:2006ee}%
  \BibitemOpen
  \bibfield  {author} {\bibinfo {author} {\bibfnamefont {H.}~\bibnamefont
  {Hansen}}, \bibinfo {author} {\bibfnamefont {W.~M.}\ \bibnamefont
  {Alberico}}, \bibinfo {author} {\bibfnamefont {A.}~\bibnamefont {Beraudo}},
  \bibinfo {author} {\bibfnamefont {A.}~\bibnamefont {Molinari}}, \bibinfo
  {author} {\bibfnamefont {M.}~\bibnamefont {Nardi}},\ and\ \bibinfo {author}
  {\bibfnamefont {C.}~\bibnamefont {Ratti}},\ }\bibfield  {title} {\bibinfo
  {title} {{Mesonic correlation functions at finite temperature and density in
  the Nambu-Jona-Lasinio model with a Polyakov loop}},\ }\href
  {https://doi.org/10.1103/PhysRevD.75.065004} {\bibfield  {journal} {\bibinfo
  {journal} {Phys. Rev. D}\ }\textbf {\bibinfo {volume} {75}},\ \bibinfo
  {pages} {065004} (\bibinfo {year} {2007})},\ \Eprint
  {https://arxiv.org/abs/hep-ph/0609116} {arXiv:hep-ph/0609116} \BibitemShut
  {NoStop}%
\bibitem [{\citenamefont {Rehberg}\ \emph {et~al.}(1996)\citenamefont
  {Rehberg}, \citenamefont {Klevansky},\ and\ \citenamefont
  {Hufner}}]{Rehberg:1995kh}%
  \BibitemOpen
  \bibfield  {author} {\bibinfo {author} {\bibfnamefont {P.}~\bibnamefont
  {Rehberg}}, \bibinfo {author} {\bibfnamefont {S.~P.}\ \bibnamefont
  {Klevansky}},\ and\ \bibinfo {author} {\bibfnamefont {J.}~\bibnamefont
  {Hufner}},\ }\bibfield  {title} {\bibinfo {title} {{Hadronization in the
  SU(3) Nambu-Jona-Lasinio model}},\ }\href
  {https://doi.org/10.1103/PhysRevC.53.410} {\bibfield  {journal} {\bibinfo
  {journal} {Phys. Rev. C}\ }\textbf {\bibinfo {volume} {53}},\ \bibinfo
  {pages} {410} (\bibinfo {year} {1996})},\ \Eprint
  {https://arxiv.org/abs/hep-ph/9506436} {arXiv:hep-ph/9506436} \BibitemShut
  {NoStop}%
\bibitem [{\citenamefont {Cleymans}\ \emph {et~al.}(2006)\citenamefont
  {Cleymans}, \citenamefont {Oeschler}, \citenamefont {Redlich},\ and\
  \citenamefont {Wheaton}}]{Cleymans:2005xv}%
  \BibitemOpen
  \bibfield  {author} {\bibinfo {author} {\bibfnamefont {J.}~\bibnamefont
  {Cleymans}}, \bibinfo {author} {\bibfnamefont {H.}~\bibnamefont {Oeschler}},
  \bibinfo {author} {\bibfnamefont {K.}~\bibnamefont {Redlich}},\ and\ \bibinfo
  {author} {\bibfnamefont {S.}~\bibnamefont {Wheaton}},\ }\bibfield  {title}
  {\bibinfo {title} {{Comparison of chemical freeze-out criteria in heavy-ion
  collisions}},\ }\href {https://doi.org/10.1103/PhysRevC.73.034905} {\bibfield
   {journal} {\bibinfo  {journal} {Phys. Rev. C}\ }\textbf {\bibinfo {volume}
  {73}},\ \bibinfo {pages} {034905} (\bibinfo {year} {2006})},\ \Eprint
  {https://arxiv.org/abs/hep-ph/0511094} {arXiv:hep-ph/0511094 [hep-ph]}
  \BibitemShut {NoStop}%
\bibitem [{\citenamefont {Abdallah}\ \emph {et~al.}(2022)\citenamefont
  {Abdallah} \emph {et~al.}}]{STAR:2021fge}%
  \BibitemOpen
  \bibfield  {author} {\bibinfo {author} {\bibfnamefont {M.~S.}\ \bibnamefont
  {Abdallah}} \emph {et~al.} (\bibinfo {collaboration} {STAR}),\ }\bibfield
  {title} {\bibinfo {title} {{Measurements of Proton High Order Cumulants in
  $\sqrt{s_{_{\mathrm{NN}}}}$ = 3 GeV Au+Au Collisions and Implications for the
  QCD Critical Point}},\ }\href
  {https://doi.org/10.1103/PhysRevLett.128.202303} {\bibfield  {journal}
  {\bibinfo  {journal} {Phys. Rev. Lett.}\ }\textbf {\bibinfo {volume} {128}},\
  \bibinfo {pages} {202303} (\bibinfo {year} {2022})},\ \Eprint
  {https://arxiv.org/abs/2112.00240} {arXiv:2112.00240 [nucl-ex]} \BibitemShut
  {NoStop}%
\bibitem [{\citenamefont {Ablyazimov}\ \emph {et~al.}(2017)\citenamefont
  {Ablyazimov} \emph {et~al.}}]{CBM:2016kpk}%
  \BibitemOpen
  \bibfield  {author} {\bibinfo {author} {\bibfnamefont {T.}~\bibnamefont
  {Ablyazimov}} \emph {et~al.} (\bibinfo {collaboration} {CBM}),\ }\bibfield
  {title} {\bibinfo {title} {{Challenges in QCD matter physics --The scientific
  programme of the Compressed Baryonic Matter experiment at FAIR}},\ }\href
  {https://doi.org/10.1140/epja/i2017-12248-y} {\bibfield  {journal} {\bibinfo
  {journal} {Eur. Phys. J. A}\ }\textbf {\bibinfo {volume} {53}},\ \bibinfo
  {pages} {60} (\bibinfo {year} {2017})},\ \Eprint
  {https://arxiv.org/abs/1607.01487} {arXiv:1607.01487 [nucl-ex]} \BibitemShut
  {NoStop}%
\bibitem [{\citenamefont {Adamczewski-Musch}\ \emph {et~al.}(2020)\citenamefont
  {Adamczewski-Musch} \emph {et~al.}}]{HADES:2020wpc}%
  \BibitemOpen
  \bibfield  {author} {\bibinfo {author} {\bibfnamefont {J.}~\bibnamefont
  {Adamczewski-Musch}} \emph {et~al.} (\bibinfo {collaboration} {HADES}),\
  }\bibfield  {title} {\bibinfo {title} {{Proton-number fluctuations in $\sqrt
  {s_{NN}}$ =2.4 GeV Au + Au collisions studied with the High-Acceptance
  DiElectron Spectrometer (HADES)}},\ }\href
  {https://doi.org/10.1103/PhysRevC.102.024914} {\bibfield  {journal} {\bibinfo
   {journal} {Phys. Rev. C}\ }\textbf {\bibinfo {volume} {102}},\ \bibinfo
  {pages} {024914} (\bibinfo {year} {2020})},\ \Eprint
  {https://arxiv.org/abs/2002.08701} {arXiv:2002.08701 [nucl-ex]} \BibitemShut
  {NoStop}%
\end{thebibliography}%

\end{document}